\documentclass[fleqn,usenatbib]{mnras}

\usepackage{graphicx}	
\usepackage{amsmath}	% Advanced maths commands
\usepackage{amssymb}
\usepackage{subcaption}
\usepackage{multirow}
\usepackage{amsmath}
\usepackage{url}
\usepackage{setspace}
\usepackage{longtable}
\usepackage{pdflscape}
\usepackage{float}
\usepackage{booktabs}
\usepackage{tabularx}
\usepackage{amssymb}
\usepackage{wasysym}
\usepackage{rotating}
\usepackage{adjustbox}
\usepackage{xspace}
\usepackage{xcolor}
\usepackage[T1]{fontenc}

\newcommand{\kms}{km s$^{-1}$}
\newcommand{\mspc}{M$_\odot$ pc$^{-2}$}

\newcommand{\hi}{{H\,{\small I}}\xspace }

\title[Resolving the Disc-Halo Degeneracy: NGC 6946]{Resolving the Disc-Halo Degeneracy II: NGC 6946}
\author[S. Aniyan, A.A. Ponomareva, K.C. Freeman et al.]
{\parbox{\textwidth}{S. Aniyan$^1$, A. A. Ponomareva$^{2,1,3}$\thanks{Email: anastasia.ponomareva@physics.ox.ac.uk}, K. C. Freeman$^1$\thanks{Email: kenneth.freeman@anu.edu.au}, M. Arnaboldi$^{4}$, O. E. Gerhard$^5$, L. Coccato$^4$, K. Kuijken$^6$ \& M. Merrifield$^7$ }\vspace{0.4cm}\\  
$^1$Research School of Astronomy \& Astrophysics, Australian National University, Canberra, ACT 2611, Australia \\
$^2$Oxford Astrophysics, Denys Wilkinson Building, University of Oxford, Keble Rd, Oxford, OX1 3RH, UK\\
$^3$Kapteyn Astronomical Institute, University of Groningen, Postbus 800, NL-9700 AV Groningen, The Netherlands \\
$^4$European Southern Observatory, Karl-Schwarzschild-Strasse 2, D-85748 Garching, Germany\\
$^5$Max-Planck-Institut f{\"u}r Extraterrestrische Physik, Giessenbachstrasse, 85741 Garching, Germany \\
$^6$Leiden Observatory, Leiden University, Niels Bohrweg 2, NL-2333 CA Leiden, the Netherlands \\
$^7$School of Physics and Astronomy, University of Nottingham, University Park, Nottingham, NG7 2RD, UK}

\date{Accepted XXX. Received YYY; in original form ZZZ}

\pubyear{2020}

\begin{document}
\label{firstpage}
\pagerange{\pageref{firstpage}--\pageref{lastpage}}
\maketitle

\begin{abstract}
The mass-to-light ratio ($M/L$) is a key parameter in decomposing galactic rotation curves into contributions from the baryonic components and the dark halo of a galaxy. One direct observational method to determine the %
disc %
$M/L$ is by calculating the surface mass density of the disc  %, which depends on 
from the stellar vertical velocity dispersion and the scale height of the disc. Usually, the scale height is obtained from near-IR studies of edge-on galaxies %% 
and pertains  %%
to the older, kinematically hotter stars in the disc, while the vertical velocity dispersion of stars is measured in the optical band and refers to stars of all ages (up to $\sim$ 10 Gyr) %
and velocity dispersions. This mismatch between the scale height and the velocity dispersion can %result 
lead to underestimates of the disc surface density and % in the 
a misleading conclusion of the sub-maximality of galaxy discs. In this paper we present the study of the stellar velocity dispersion of the 
% face-on 
disc galaxy NGC~6946 using integrated star light %as well as 
and individual planetary nebulae as dynamical tracers. %
We demonstrate the presence of two kinematically distinct populations of tracers which contribute to the total stellar velocity dispersion. Thus, we %
are able to use the dispersion and the scale height of the same % 
dynamical %
population to derive %their 
the surface mass density %
of the disc %
over a radial extent. % As a result, 
We find the disc of NGC~6946 to be closer to maximal with the baryonic component contributing  
most of the radial gravitational field in the inner parts of the galaxy  ($\rm V_{max}(bar)=0.76(\pm0.14)V_{max}$).

\end{abstract}


\defcitealias{Aniyan:18}{An18}

\section{Introduction}
The "disc-halo" degeneracy is an important issue when decomposing \hi rotation curves of galaxies, where the contribution from the baryonic matter is mostly determined by the stellar mass-to-light ratio ($M/L$). Unfortunately, the $M/L$ is %very 
uncertain due to the challenges involved in traditional methods used to determine it \citep{vanAlbada, Maraston:2005, Conroy:2009}. Accurate rotation curve decomposition is crucial in determining the potential of a galaxy % including 
and the parameters of %the 
its dark matter (DM) halo. The densities and scale radii of dark haloes are known to follow well-defined scaling laws and can therefore be used to measure the redshift of assembly of haloes of different masses \citep{Maccio:13, Kormendy:16, somerville18}.

A rather direct observational technique of measuring the %
disc $M/L$ is from its surface mass density, which can be estimated from the vertical velocity dispersion and the vertical scale height of the disc \citep{VanFree, Bottema:87, HC1, DMI}. The 1D Jeans equation in the vertical direction can be used to estimated the surface mass density ($\Sigma$) of the disc via the relation:
\begin{equation}
 \rm \Sigma = f\sigma_z^2/Gh_z
 \label{main_eq}
\end{equation}
where $h_z$ is the vertical scale height and $\sigma_z$ is the %
integrated vertical velocity dispersion of the exponential disc, $\rm G$ is the gravitational constant and $f$ is a geometric factor, known as the vertical structure constant, that depends weakly on the adopted vertical structure of the disc.
Having estimated $\Sigma$, we can infer stellar mass surface density ($\Sigma_{\star}$) and the $M/L$ as $\Upsilon_{\star} =\Sigma_{\star}/\mu$, where $\mu$ is the surface brightness of a galaxy in physical units ($\rm L_{\odot} %
~pc^2$), thus breaking the disc-halo degeneracy.

In \citealt{Aniyan:18} (hereafter \citetalias{Aniyan:18}), we argued that for the Jeans equation (Eqn. \ref{main_eq}) to work effectively, the scale height and dispersions must refer to the same population of stars, and in this case the $\sigma_z$ is expected to fall exponentially with twice the galaxy scale length. However, usually the vertical dispersions are obtained from measurements of near-face-on galaxies in the optical bands, whereas the scale heights are obtained from studies of edge-on galaxies in the red and NIR bands \citep{Kregel:02}\footnote{ See also \citet{Kregel:05} for a detailed study of the correlation between the intrinsic properties of edge-on galaxies, i.e. central surface brightness, scale length and scale height in the I-band.}.Thus, the spectra of star-forming face-on disc galaxies include light from the younger, kinematically colder population of stars. On the other hand, the scale heights obtained from observations of edge-on discs are primarily for the kinematically hotter stars which are above the dust lane in the galaxies. This mismatch can lead to the surface mass density being underestimated which can in turn lead to galaxies' discs being incorrectly classified as sub-maximal. This issue has plagued most of the previous measurements of the surface mass density of face-on discs \citep{HC3, DMS5}.

In \citetalias{Aniyan:18} we used integrated light data as well as planetary nebulae (PNe) as tracers of the kinematics of the disc in the face-on spiral NGC 628 to demonstrate the presence of two kinematically distinct populations of disc stars in this galaxy. We also showed the effect of the cold layer on the total surface mass density of the disc (see Appendix in \citetalias{Aniyan:18}). This is based on an exact solution for an isothermal sheet of older stars with an embedded very thin layer of younger stars and gas. Its density distribution is a modified version of the familiar ${\rm sech}^{2}(z/2h_z)$ distribution. Following this solution equation \ref{main_eq} becomes:
\begin{equation}
\Sigma_T = \Sigma_D + \Sigma_{C,*} + \Sigma_{C,gas} = \sigma_z^2/(2 \pi G h_z),
\label{main_eq2}
\end{equation}
where $\Sigma_T$ is the total surface density of the disc and $\Sigma_D$ is the surface density of the older stellar component which is used as the dynamical tracer (its scale height is $h_z$ and its isothermal vertical velocity dispersion is $\sigma_z$).  $\Sigma_{C,*}$ and $\Sigma_{C,gas}$ are the surface densities of the cold thin layers of young stars and gas respectively. An independent measurement of $\Sigma_{C,gas}$ is available from 21-cm and mm radio observations. As a result, in \citetalias{Aniyan:18} we showed that the disc which was primarily considered sub-maximal \citep{HC3} appears maximal with $V_{\rm baryonic} = (0.78 \pm 0.11)V_{\rm max}$.

 In this paper, we extend our previous work to describe the kinematics of the nearby disc galaxy NGC~6946 with the purpose to establish the method's general viability. The choice of NGC~6946 is motived by its inclination ($\sim$ 37$^\circ$), as it is not as face-on as NGC~628, which allows us to derive a more reliable rotation curve, as well as makes more targets accessible for the analysis. We observe NGC~6946 with the VIRUS-W integral field unit spectrograph in the inner regions (30-125'') to study stellar kinematics through the absorption lines, and with the Planetary Nebula Spectrograph (PN.S), which allows us to study the kinematics of the stellar component in galaxies at low surface brightness values in ellipticals \citep{pulsoni18}, lenticulars \citep{cortesi13} and discs \citepalias{Aniyan:18} using PNe as discrete tracers out to large radii. The use of PNe to map kinematics to large radii in discs showed the presence of a cold younger disc and a older hotter thicker disc in the Andromeda galaxy in the radial range 14-28 kpc  \citep{Bhattacharya:19}. NGC~6946 is a nearby (D = 6.1 Mpc, $1'' = 29.6$ pc) disc galaxy with a high star formation rate (SFR $=$ 4.8 M$_\odot$ yr $^{-1}$ \citealp{Lee:06}), which is $\sim$ 4 times the SFR of NGC~628. This contributes to the importance of the NGC~6946 analysis to access differences between the systems. Unfortunately, the inclination of this galaxy brings in some challenge, as it gets harder to separate the in-plane dispersion components ($\sigma_R$ and $\sigma_\phi$) from the vertical dispersion ($\sigma_z$). However, NGC~6946 is closer than NGC~628, and thus we can achieve higher S/N data for the PNe which helps in our analysis. With the PN.S., PNe in NGC 6946 can be detected and their velocities measured out to $388'' = 11.5$ kpc (equivalent to 4.1 disc scale lengths) reaching a surface brightness value of 25 mag in B-band. 

This paper is organised as follows: Section 2 describes the observations and data reduction for VIRUS-W. Section 3 summarises the same for the PN.S. Section 4 discusses the photometric properties and derives scale height of the galaxies. Section 5 discusses our analysis to derive the surface mass density of the cold gas in this galaxy. Section 6 presents the analysis involved in the extraction of a double Gaussian model from our data. Section 7 discusses the vertical dispersion profile of the hot and cold stellar components. Section 8 describes the calculation of the stellar surface mass density. Section 9 explains the rotation curve decomposition using the calculated surface mass densities. Section 10 presents our conclusions and scope for future work.

\section{VIRUS-W Spectrograph}
\subsection{Observations}
\begin{figure*}%fig 1
 \includegraphics[width=1\linewidth]{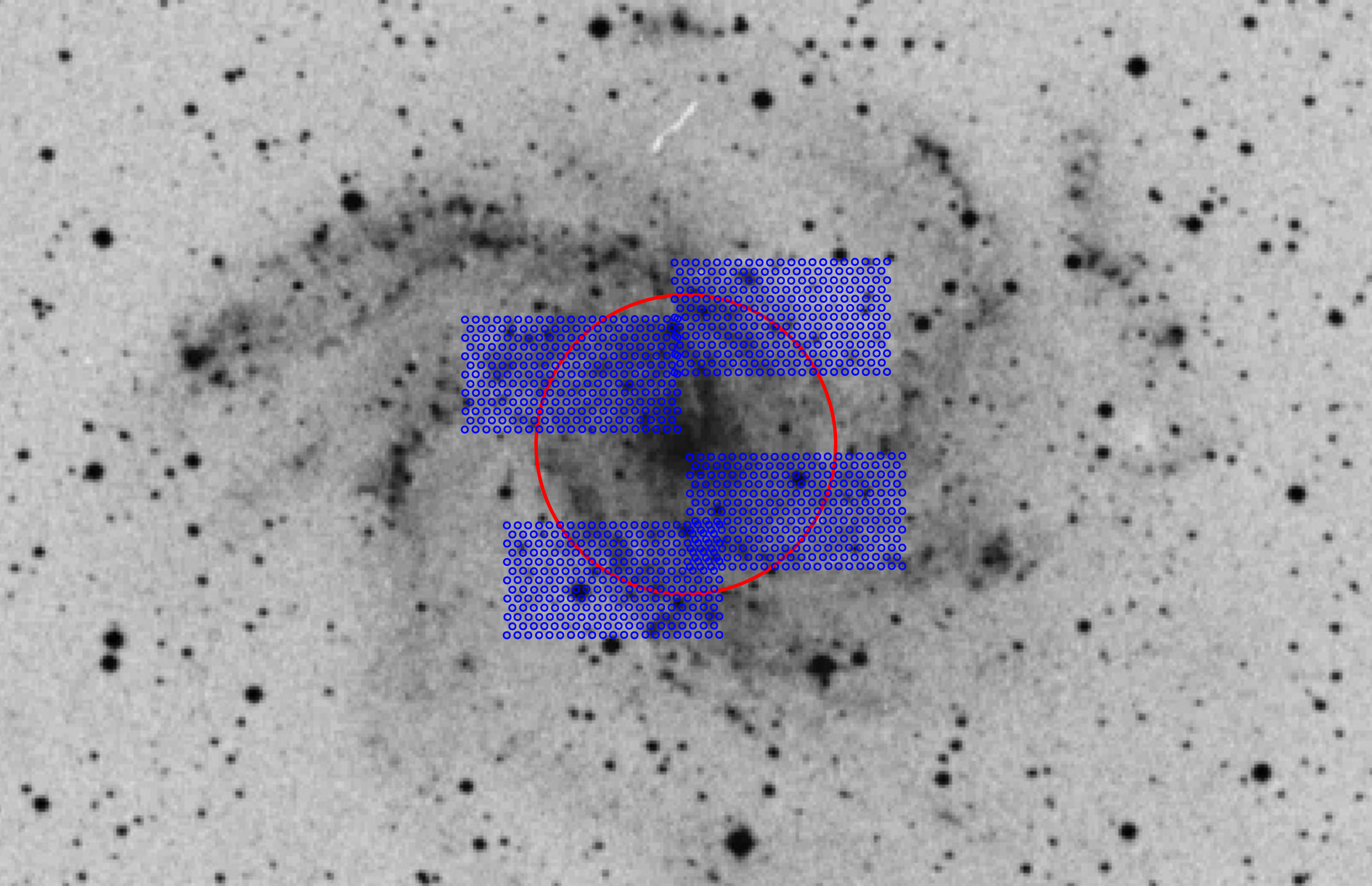}
\caption{The positions of the four VIRUS-W IFU fields % together with 
showing the 267 fibres in each field overlaid on a DSS image of NGC 6946. The red circle at %
a radius of $75''$ shows % where we separated 
the separation between our %data into the 
inner and outer radial bins.}
\label{VW_field_6946}	
\end{figure*}

VIRUS-W is an IFU spectrograph on the 2.7-m telescope at McDonald Observatory designed for relatively high resolution spectroscopy of low surface brightness regions of galaxies \citep{fabricus12}. VIRUS-W observations for NGC 6946 were carried out in October 2014. We were able to get good quality data for 4 fields which were positioned on the galaxy at a luminosity weighted radius of about 1 radial scale length along the major and minor axis, and cover the radial extend of up to 175'' which corresponds to $\approx 20.6$ $\rm mag/arcsec^{2}$ in I-band, and the surface brightness in the V-band
(in which the Virus W observations were made) is about a magnitude fainter than the dark V-band sky (see Figure \ref{fig_SBPS}). The positions of the IFU fields are shown in Figure \ref{VW_field_6946}.
The coordinates and exposure time at each position are given in Table \ref{VW_pos_6946}. The observations were carried out %through 
in a sky -- galaxy -- sky observing sequence. This sequence was repeated at least % thrice 
three times at each field, as indicated in Table \ref{VW_pos_6946}. This enabled very good sky subtraction using the automated pipeline developed for VIRUS-W (see \citetalias{Aniyan:18} for further details).
\begin{table} %table 1
\centering
\begin{tabular}{c c c c}
\hline 
Galaxy & RA (J2000) & Dec (J2000) & Exposure Time (s) \\
\hline
\multirow{4}{*}{NGC 6946} & 20:34:54.04                                    & +60:08:04.8                                      & 3 $\times$ 800                                                 \\
& 20:34:47.25                                      & +60:10:23.8                                    & 3 $\times$ 800                                                 \\
& 20:34:47.13                                      & +60:08:48.4                                       & 3 $\times$ 800                                                 \\
& 20:34:56.34                                    & +60:10:01.6                                     & 3 $\times$ 800                                                 \\     
\hline 
\end{tabular}
\caption{Coordinates and exposure times for the IFU fields.}
\label{VW_pos_6946}
\end{table}

% Further, 
We then used the CURE data reduction pipeline \citep{cure} to get the sky subtracted 1D % spectrum 
spectra at each of the four fields. Since the IFU fields cover a large radial extent of NGC~6946 (Figure \ref{VW_field_6946}), we split the data into two radial bins at luminosity-weighted mean radii of $54''$ and $98''$ respectively. After further processing (see Section \ref{sec_sas}), we obtain a 1D summed spectrum for each radial bin that we use to extract the velocity dispersions from the absorption lines.

\section{Planetary Nebula Spectrograph}
\subsection{Observations and Velocity Extraction}  %\S3.1
The PN.S is a double counter-dispersed wide-field spectrograph designed to discover extragalactic planetary nebulae and measure their radial velocities in a single observation, used on the William Herschel Telescope on La Palma \citep{douglas02}. 
The data for NGC 6946 were acquired in September 2014. The weather during the run was excellent, with typical seeing of $\sim 1''$.  We obtained 11 images centred on the centre of the galaxy, each with an exposure time of 1800s following the strategy adopted in \citetalias{Aniyan:18}. We use the PN.S data reduction pipeline \citep{Douglas:07} to get the final stacked 'left' and 'right' image for the 
$\rm [OIII]$ data. % Identifying 
The unresolved objects are candidate planetary nebulae, and their relative positions on the left and right stacked images give us their radial velocities. % of these sources.  
We also imaged NGC 6946 in H$\alpha$, using the H$\alpha$  narrow band filter on the undispersed H$\alpha$ arm of the PN.S. We then used the [OIII] stacked images along with the H$\alpha$ stacked image to identify the PNe candidates in this galaxy. %

\subsection{Identification of Sources}
\begin{figure} %fig 2
\centering
   \includegraphics[width=\linewidth]{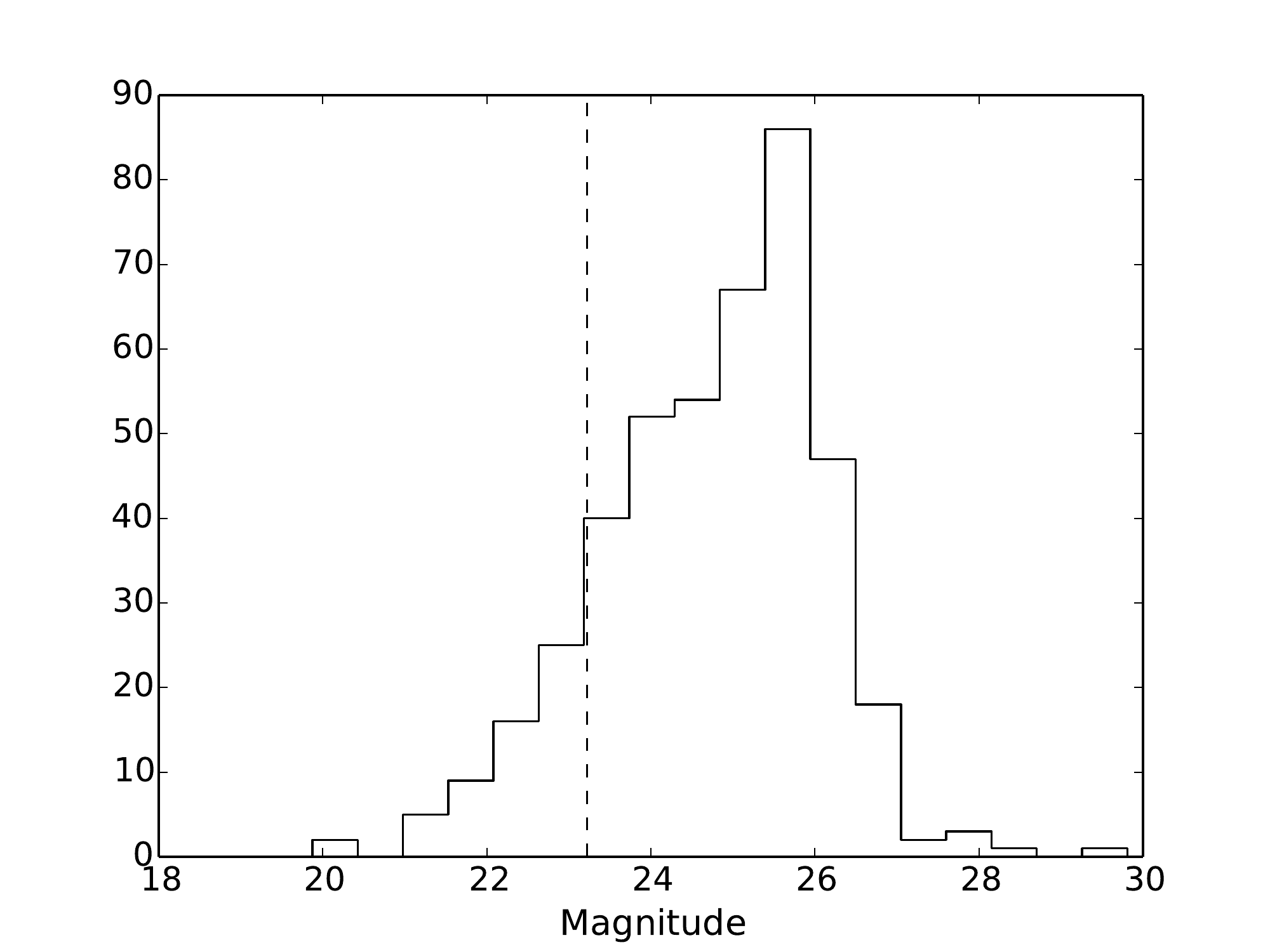}
   \caption[The luminosity function for NGC 6946]{The luminosity function for all spatially unresolved [OIII] emitters identified in the combined left and right images of the PN.S for NGC 6946. The dashed line shows the expected bright luminosity cut-off for PNe. Objects brighter than the cut-off are % most 
probably bright HII regions.}
\label{Lum_fun_N6946}
\end{figure}

To identify our  % true 
PNe and separate them from the % H2 
HII regions which also emit in O[III], we use the luminosity function for all spatially unresolved [OIII] emitters identified in the combined left and right images of the PN.S. Then we introduce the expected bright luminosity cut-off for PNe. We include only objects fainter than this value in our analysis. Objects brighter than the cut-off are mostly obvious bright HII regions. For more details please see \citetalias{Aniyan:18}. Consequently we identified 444 unresolved objects with [OIII] emission in this galaxy. From the measured positions of these sources on the left and right images, astrometric positions and line-of-sight (LOS) velocities were derived simultaneously. %
The typical measurement error associated with the PNe radial velocities is $<$~9 \kms\ (see Section 3.3 in \citetalias{Aniyan:18}). 
% Following, 
We %
then converted our instrumental magnitudes % into
on to the $m_{5007}$ magnitude scale $m_{5007} = m_{0} + 25.16$, using our spectrophotometric standards.  % Then, we corrected 
These magnitudes %
were corrected for foreground extinction using %
the \citet{Schlafly:2011} dust maps. % The 
At the distance of NGC 6946, the bright luminosity cut-off for PNe in this galaxy is expected to be at $m_{5007} = 23.22$. Figure \ref{Lum_fun_N6946} shows the luminosity function for NGC 6946, including all identified unresolved sources. The dashed line in the plot shows the position of the bright luminosity cut-off for the PNe. 

\begin{figure} %fig 3
\centering
 \includegraphics[width=\linewidth]{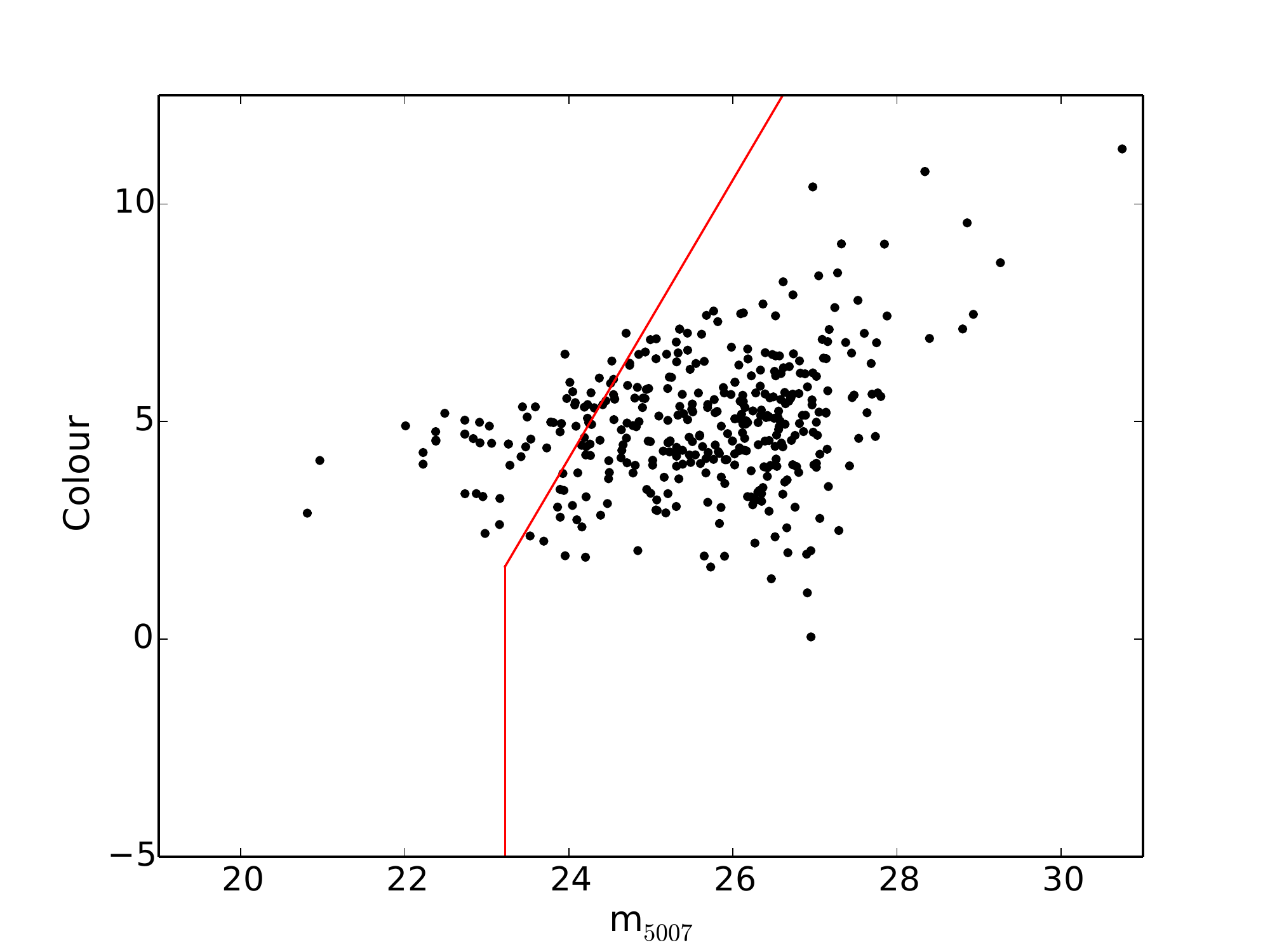}
 \caption[Colour-Magnitude plot for NGC 6946]{The colour-magnitude plot for
 [OIII] sources in NGC 6946. The vertical axis is the ([OIII] - H$\alpha$) colour and the horizontal axis is the m$_{5007}$ magnitude. The PNe lie to the right of the red lines. Only these objects are used in our analysis.}
\label{CMD_N6946}
\end{figure}

After the identification of 444 sources, the resulting sample represents a mix of HII regions and PNe since both can have strong [OIII] emission. We again use our empirical relationship between the ([OIII] -- H$\alpha$) colour and the m$_{5007}$ magnitude to discriminate between PNe and HII regions \citep{XXXX:17}. Figure \ref{CMD_N6946} shows the colour-magnitude diagram used to identify PNe. Thus, we obtain 375 unresolved [OIII] sources classified as PNe: 125 per each radial bin. Earlier observations by \citet{HC2}  found $\approx$ 70 PNe candidates in this galaxy.

In addition to HII regions, the historical supernovae are a potential source of contamination which can bias the planetary nebulae luminosity function \citep{kreckel17}. According to \href{http://www.cbat.eps.harvard.edu/lists/Supernovae.html}{the IAU Central Bureau for Astronomical Telegrams (CBAT) List of Supernovae}, there are nine known historical supernovae in NGC 6946, which is an unusually high number. However, none of these objects made it into our PNe sample. We had one unresolved [OIII] source at $\sim$ $3''$ from the historical supernova SN 2002hh. Yet, this object was classified as an HII region after applying our colour-magnitude cut. Thus, we don't have any of these contaminants in our final PNe sample.  %

\section{Photometry and Scale Height}  % \S 4
\label{sec_photometry}
\begin{figure} %fig 4
\centering
 \includegraphics[width=\linewidth]{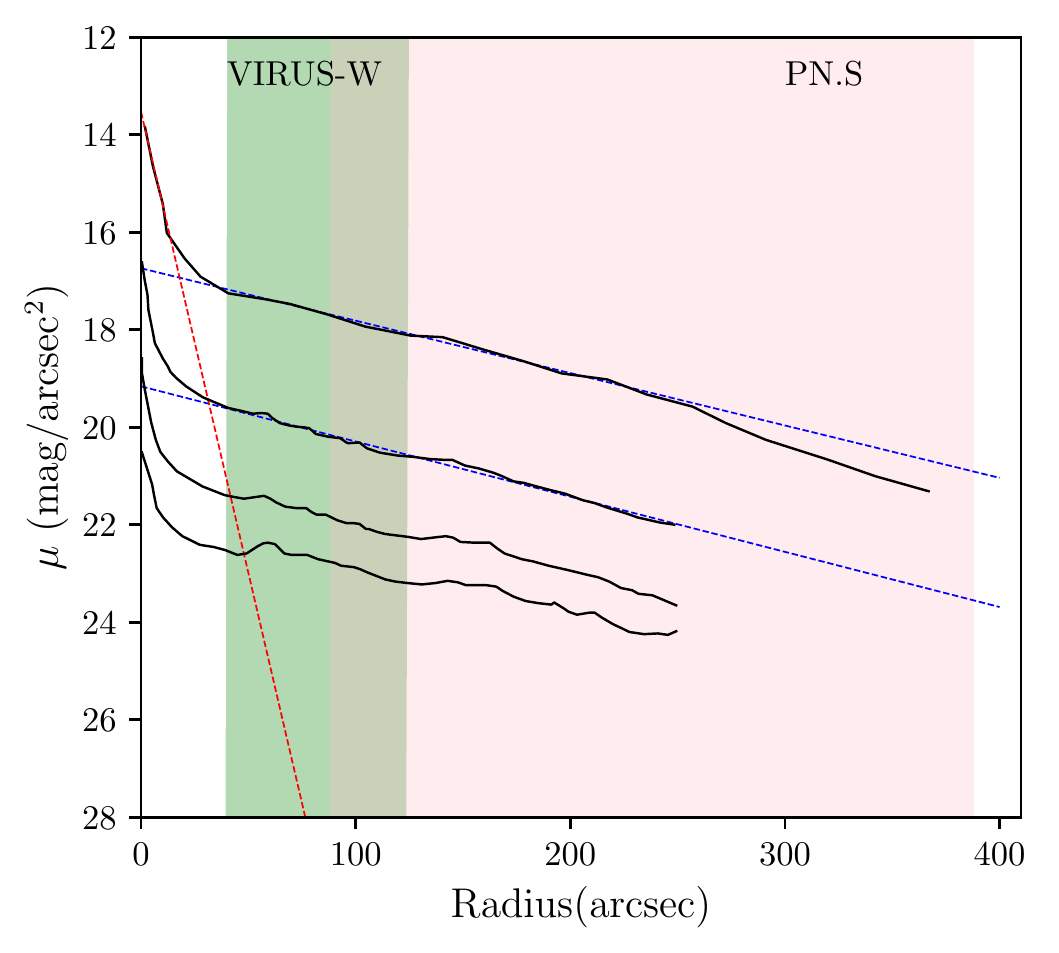}
 \caption{The surface brightness profiles of NGC 6946 in BVI \citep{Makarova:99} and 3.6 $\rm \mu$m band \citep{munoz:09} are shown from bottom to top. The surface brightness is shown in VEGA magnitudes. The contribution to the light from the stellar disc in I-band is shown with blue line. The contributions to the total light from bulge and disc in the 3.6 $\rm \mu$m band are shown with red and blue lines respectively. The average error on the surface brightness is $\sim$0.01 mag for the 3.6 $\rm \mu$m band and $\sim$0.02 mag in the optical bands. The radial extent of the Virus-W and PNS data is shown with green and pink bands respectively.}
\label{fig_SBPS}
\end{figure}
NGC~6946 is a late-type spiral galaxy which shows the presence of a small bulge. Figure \ref{fig_SBPS} shows the surface brightness profiles of NGC~6946 in four photometrical bands: BVI from \citet{Makarova:99} and 3.6 $\rm \mu$m from \citet{munoz:09}. We use spatial information from these profiles % further 
later in the paper to calculate the mass-to-light ratio as a function of radius. We also use the 3.6 $\rm \mu$m profile to perform bulge-disc decomposition (shown with red and blue lines in Figure \ref{fig_SBPS}) in order to account for the bulge contribution during the mass modelling. 

The disc scale height ($\rm h_{z}$) cannot be directly measured for face-on galaxies. Thus, the scaling relation between the scale height and scale length ($\rm h_{R}$) of spiral galaxies is often used to infer $\rm h_{z}$ of a face-on system. \citet{Kregel:02} studied the scale heights of a sample of edge-on galaxies in the I-band and found strong correlations with the scale lengths. The use of the I-band is justified as it is not sensitive to the contribution of dust and PAHs, % as well as 
and it traces the older thicker stellar disc population. % residing in a thicker disc. 
\citet{Bershady:10b} found that $\log(h_R / h_z) = 0.367\log(h_R/{\rm kpc}) + 0.708 \pm 0.095$ by fitting the data from \citet{Kregel:02}.\footnote{\citet{Bershady:10b} scaling is used to convert the scale length into scale height for an old population of stars.} Thus, using the disc fit to the I-band surface brightness profile (Figure \ref{fig_SBPS}) we measure the disc scale length to be equal to 95" or 2.8 kpc. Throughout the paper we adopt %
the distance to the galaxy D = $6.1 \pm 0.6$ Mpc from \citet{HC1} to be consistent with our previous study of NGC 628 \citepalias{Aniyan:18}.
We note that this distance is a bit smaller than the distance determined by \citet{anand18} ($7.72 \pm 0.32$ Mpc) who used the tip of the giant branch method. Finally, we obtain the scale height of the disc $h_z = 376 \pm 75$ pc. We use this value throughout the paper when we refer to the calculation of the surface mass density and mass modelling. Following \citet{degrijs97} we assume that the scale height $h_z$ for the disc of this late-type spiral is independent of the galactic radius. 

Table \ref{par_N6946} summarises the values of the various parameters of this galaxy that we use in our analysis of the surface mass density and mass modelling. 
Inclination and position angles (PA) are measured with the tilted-ring modelling of the THINGS HI data (\citealt{things}, see Section \ref{sec_rotcurve}).
These values  agree well with the values obtained from the PNe. This indicates that there is no evident disconnect between the gaseous HI disc and the stellar disc, at least within our covered radii.

\begin{table} % table 2
%\centering
%\resizebox{\columnwidth}{!}{
\begin{tabular}{l c }
\hline 
 Parameters & Value  \\
\hline
Scale length (I - band)$^a$ & $95.0$" $\pm 5$" \\
Scale length (3.6 $\rm \mu$m band)$^b$ & $92.0$" $\pm 1$"  \\
$\mu_{0}$[I] & 19.16 mag/arcsec$^{2}$ \\
$\mu_{0}$[3.6] & 16.74 mag/arcsec$^{2}$ \\
Scale height (I - band)& $376 \pm 75$ pc  \\
Distance$^c$ & 6.1 $\pm$ 0.6 Mpc \\
HI inclination$^d$ & 37$^\circ$ $\pm 7^\circ$ \\
HI position angle$^d$ & 242$^\circ$ $\pm 10^\circ$ \\
\hline
\end{tabular}
\caption{Photometrical and geometrical parameters of NGC 6946 measured from the literature data and adopted in the analysis throughout the paper with the associated error. The average error on the surface brightness is $\sim$0.01 mag for the 3.6 $\rm \mu$m band and $\sim$0.02 mag for the I-band. Data from the literature:
a-- \citet{Makarova:99}; b-- \citet{munoz:09}; c-- \citet{HC1}; d-- \citet{things}. 
 }
\label{par_N6946}
\end{table}

\section{Surface Mass Densities of Cold Gas}  % \S 5
To account for the gas contribution % in 
to the measured total surface mass density, and to able to separate stellar from gaseous components, we derive total cold gas surface 
density using HI data from the THINGS survey \citep{things} and CO data from the HERACLES survey \citep{heracles}. 

The HI radial surface density profile was derived from the integrated column-density HI map, constructed by summing the primary beam corrected channels of the clean data cube. 
We use the same radial sampling, position and inclination as for the tilted ring modelling (Section \ref{sec_rotcurve}). The resulting flux (Jy/beam) was converted to mass densities ($\rm M_{\odot}/\rm pc^2$) using Eqn. 5 and 6 in \citet{ponomareva16}. The error on the HI surface mass density was determined as the difference in the profile between approaching and receding sides of the galaxy, and does not exceed $\sim 0.4~$$\rm M_{\odot}/\rm pc^2$.

The H$_2$ surface mass density profile was obtained from the CO total intensity map with the same radial sampling, position and inclination angles as for the HI profile. The resulting CO intensities were converted into the H$_2$ surface mass density using the prescription by \citet{heracles}. The error on the H$_2$ surface mass density was obtained from the % the provided 
HERACLES error maps and does not exceed $\sim 0.7~$\rm M$_{\odot}/\rm pc^2$. 

\begin{figure} 
\centering
   \includegraphics[width=\linewidth]{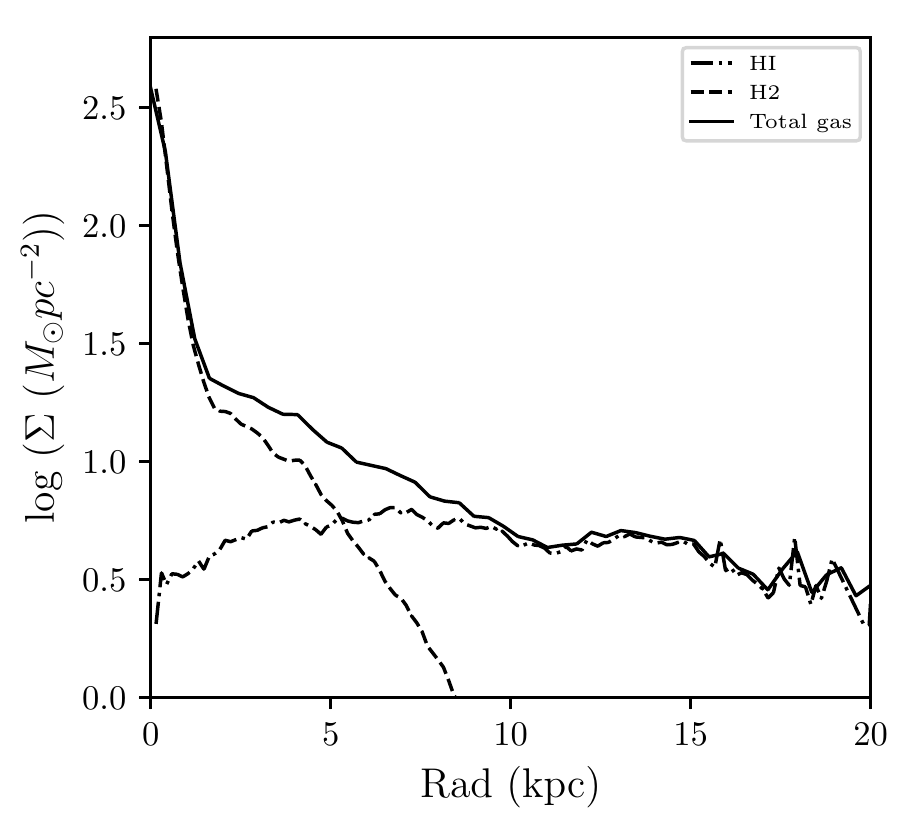}
   \caption{Surface mass density of the cold gas (atomic and molecular) in NGC 6946. The HI density profile derived from THINGS data \citep{things} is shown with the  
  dot dashed line. H2 profile derived from the HERACLES data \citep{heracles} is shown as the long-dashed curve. The surface density profile of the total gas is shown as the solid curve. All profiles are corrected for contribution from helium and heavier elements by a factor of 1.4.
}
\label{cold_gas}
\end{figure}

The resulting HI and H$_2$ surface mass density profiles together with the total cold gas profile are shown in Figure \ref{cold_gas} with dot dashed, long-dashed and solid curves respectively. All profiles are corrected for the presence of metals and helium and de-projected so as to be face-on. It is worth %mention a 
mentioning the large amount of % the 
molecular gas in NGC~6946, 
which significantly dominates the amount of atomic gas in the inner parts. % Remarkably 
We note that \citet{crosthwaite07} find similar results and report the total H$_2$ mass to be equal to $3\times10^9$ $\rm M_{\odot}$, approximately one-third of the total interstellar hydrogen gas mass. 

\begin{figure} %fig 5
\includegraphics[width=1\linewidth]{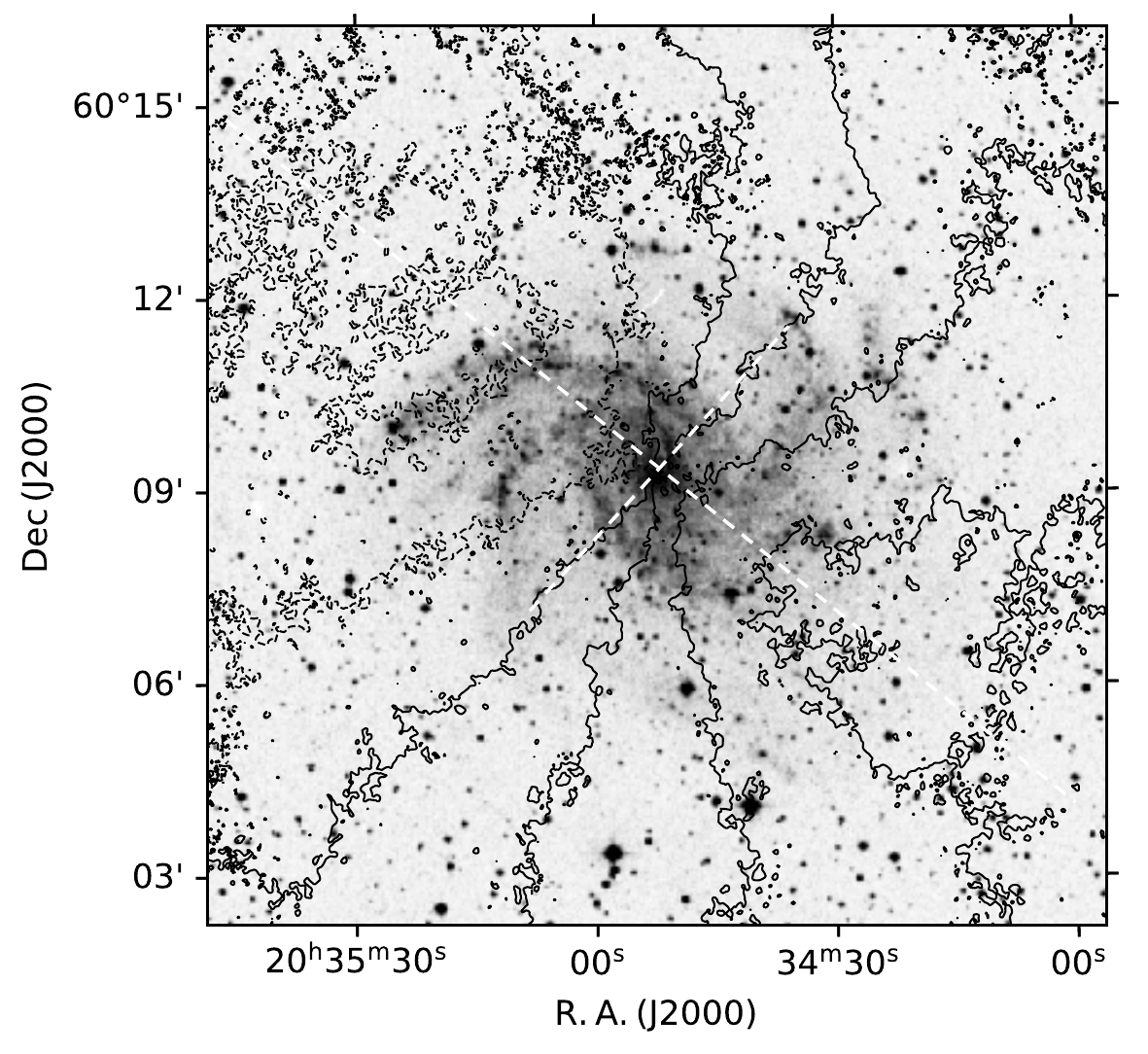}
\caption{The DSS image of NGC 6946 with the overlapped velocity contours from the HI velocity field. The orientation of the major and minor axis of the HI disc are shown with the dashed lines. Velocity separation of the contours is 40 kms$^{-1}$.}
\label{DSS_HI}
\end{figure}

\section{Extracting Velocity Dispersions of the Hot and Cold Components}

\begin{figure*}%fig 6
\centering
   \begin{subfigure}{0.8\textwidth}
	\centering
   \includegraphics[width=1\linewidth]{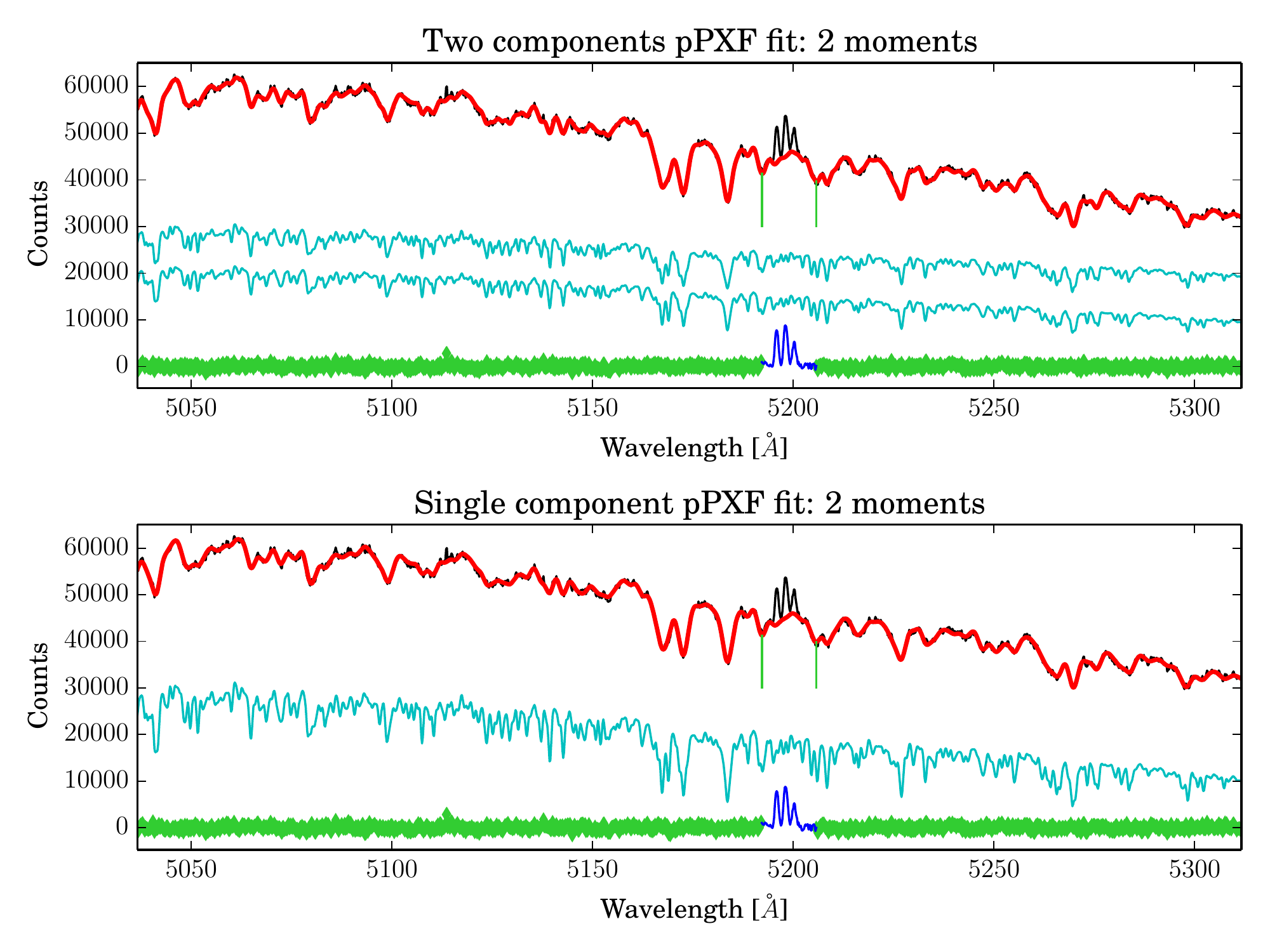}
   \caption{}
\end{subfigure}
\begin{subfigure}{0.8\textwidth}
	\centering
   \includegraphics[width=1\linewidth]{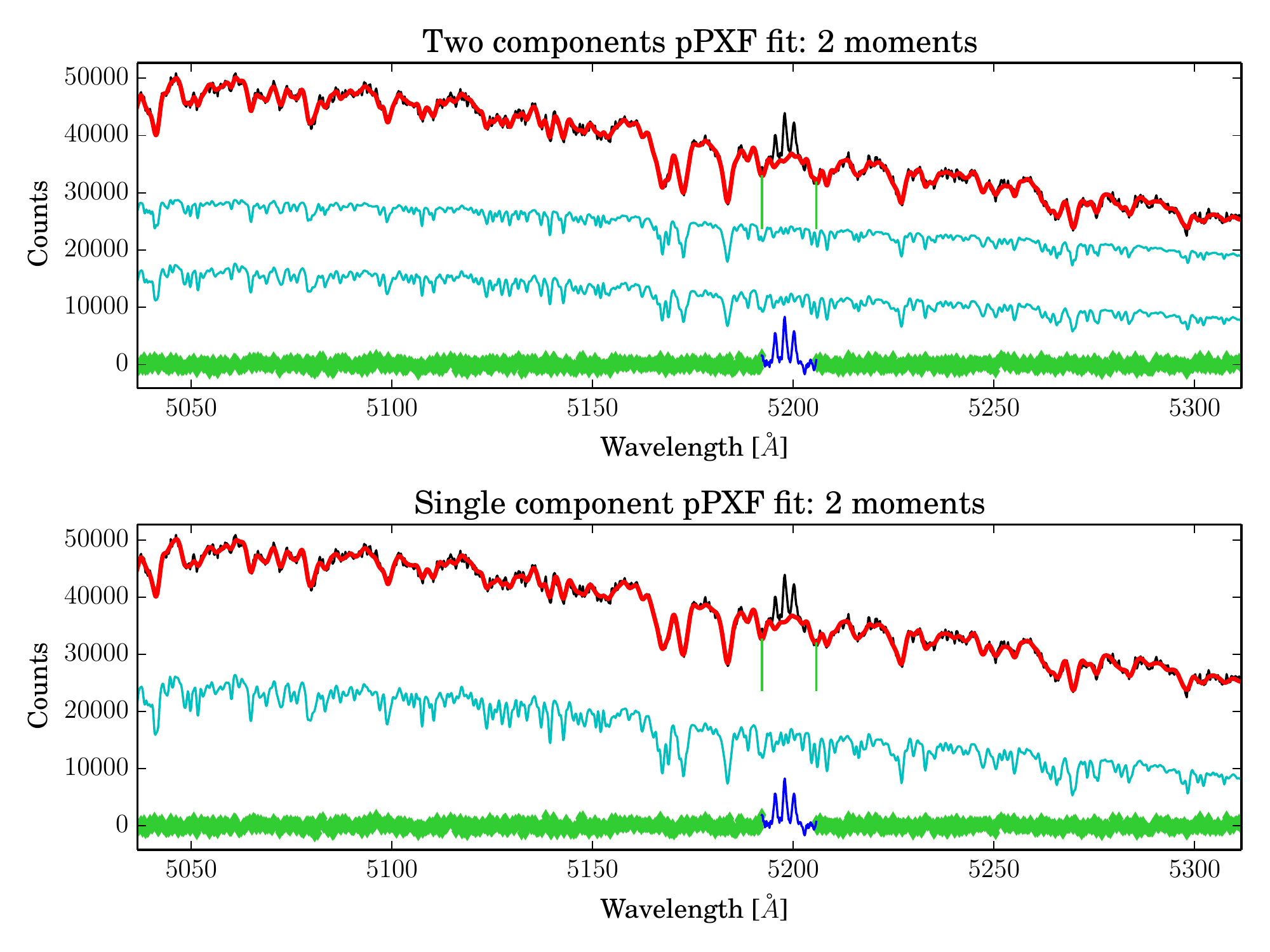}
   \caption{}
\end{subfigure}
\caption[The pPXF best fit spectra for NGC 6946.]{The pPXF fit results in (a) the inner radial bin at a luminosity weighted % distance 
mean radius of $54''$ and (b) the outer bin at a luminosity weighted % distance 
mean radius of $98''$. The upper panel shows the 2 component fit to the data whereas the lower panel shows a single component fit. Only the high resolution Mgb region of the spectrum was used for the fit. The galaxy spectrum is in black and the best fit from pPXF is in red. The cyan spectra are the two and one component spectra that pPXF found. The cyan spectra have been shifted vertically so as to be clearly visible. The residuals are shown in green. We note the presence of three peaks in the wavelength range 5196.2 - 5200.4 \AA \ which we suggest are due to the superposition of the [NI] doublet emissions from outflows in the ISM \citep{boomsma08} to the [NI] doublet emission from the regular rotating ISM, see Section \ref{sec_sas1a} for additional details. The wavelength region 5196.2 - 5200.4 \AA \ is omitted from the fit.}
\label{spec_N6946}
\end{figure*}

In this section we describe in detail the procedure of extracting accurate velocity dispersions of two stellar components. We present % 
the different techniques used to derive dispersions from the stellar absorption line spectra in the inner parts, % as well as 
and from the PNe data in the outer regions. 

\subsection{Stellar Absorption Spectra}
\label{sec_sas}
\subsubsection{Removing Galactic rotation}
\label{sec_sas1a}

To remove the galactic rotation and to get rid of any large scale streaming motions across the IFU, we measure the local HI velocity at the position of each fibre from the THINGS data \citep{things} and shift each fibre spectrum by this local HI velocity. Figure \ref{DSS_HI} shows the DSS image of NGC 6946 with the overlapped velocity contours from the HI velocity field. We note that our IFU observations lie within the stellar disc of the galaxy, as shown in Figure \ref{VW_field_6946}, while the HI disc is more extended.
This method proved to be preferable to removing the galactic rotation by modelling the rotation field over the entire IFU using the observed rotation curve, as the latter gives larger line-of-sight velocity dispersions ($\sigma_{LOS}$). 
However, our technique introduces an additional small velocity dispersion component from %
the HI itself, for which a correction is needed (see Section \ref{sec_sas1}).

The VIRUS-W % IFUs cover 
IFU covers a large radial extent of the galaxy. Since the vertical velocity dispersion ($\sigma_z$) is expected to fall exponentially with twice the galaxy's scale length, we don't want each radial bin to be too large.  We divide our IFU data into two radial bins corresponding to % a 
luminosity weighted mean radii of $54''$ and $98''$. The stellar component tends to rotate more slowly than the gaseous component and neglecting this effect can lead to overestimated $\sigma_{LOS}$. NGC 6946 has an inclination of $37^{\circ}$, so this asymmetric drift may affect the measurement significantly. 
To remove the effects of differential asymmetric drift, % several small changes made here
we split the 267 fibres in each IFU field into a grid of six cells, each with about 44 fibres. 
The gradient in asymmetric drift is negligible across the small area of a cell. We sum the spectra from all fibres in a cell and cross-correlate five of the summed spectra against the sixth. This gives the shift in velocity to apply to the five spectra to match the spectrum from the sixth grid. This procedure gets rid of any differential asymmetric drift across the IFU fields. We repeat this exercise for all four IFU fields. The cross-correlated, asymmetric-drift-corrected spectra were then shifted to redshift $z = 0$, before summing up to the final spectra.
This approach has its caveats, as the hot and cold disc components will have slightly different asymmetric drifts. The difference in asymmetric drift between the hot and cold components is 5 kms$^{-1}$ evaluated at the mean radius for the inner VIRUS-W region and 6 kms$^{-1}$ for the outer region.  At 37$^\circ$ inclination, these differences become 3 kms$^{-1}$ and 4 kms$^{-1}$ respectively. These values are uncertain because of the observation errors in the dispersion, but seem unlikely to make a significant contribution to the measured velocity dispersions.
Although this effect is expected to be minor for a galaxy with an inclination of 37$^\circ$, it could be an issue for more inclined galaxies.

The final summed spectra from the inner and outer radial bins respectively have an SNR of 105 and 77 per wavelength pixel % for the spectrum 
 (each wavelength pixel is $\sim$ 0.19 \AA;  the resolving power $R = 8700$, so the Gaussian $\sigma$ of the PSF is $14.7$ \kms). We only use the region between wavelengths of about 5050 -- 5300 \AA \ in our analysis, since it has the highest resolution and avoids the emission lines at 
lower wavelengths. The [NI] doublet emission lines together with the third peak can be seen at $\sim$ 5200 \AA \ (see Figure \ref{spec_N6946}). 
Although we exclude the emission line region from our velocity dispersion fits, the presence of these three lines is unexpected. 
It is hard to definitely conclude that they come from the galaxy interstellar medium (ISM), because the heliocentric radial velocity of 
NGC 6946 is only 40 km s$^{-1}$.
In \citetalias{Aniyan:18}, NGC 628 (radial velocity 657 km s$^{-1}$) showed the two [NI] 
emission lines at $\lambda = 5197.9$ and $\lambda = 5200.3$ \AA, and they were clearly at the 
velocity of the galaxy. The strengths of the two lines are approximately equal.  
The sky [NI] lines, which typically have line ratios 5198/5200 $\sim$ 1.7, have been 
successfully subtracted in the reduction pipeline and do not appear.
In NGC 6946, our strategy for removing galactic rotation means that all lines 
from the galactic plane have velocities near zero, relative to the systemic velocity 
of the galaxy. We see three emission lines at observed $\lambda$ 5196.2, 5198.4, 
5200.4 \AA \ . Their relative line strengths can be seen in Figure \ref{spec_N6946}. 
NGC 6946 is known to have a significant amount of halo HI at velocities up to $\pm 100$ km s$^{-1}$, 
in addition to its planar HI \citep{boomsma08}\footnote{We thank the anonymous referee for pointing out the presence of the "beard" in this galaxy, which is usually associated with the extraplanar gas.}. Thus, we speculate that the three 
observed lines come from a superposition of two pairs of [NI] lines: a  weaker 
pair at near-systemic velocity from the planar gas, and a stronger pair blue shifted 
by about 2 \AA \ associated with the negative velocity gas. The superposition of 
the two pairs gives three lines as observed, with their apparent line ratios.

\begin{figure} %fig 5
\centering
   \begin{subfigure}[b]{0.5\textwidth}
   \includegraphics[width=1\linewidth]{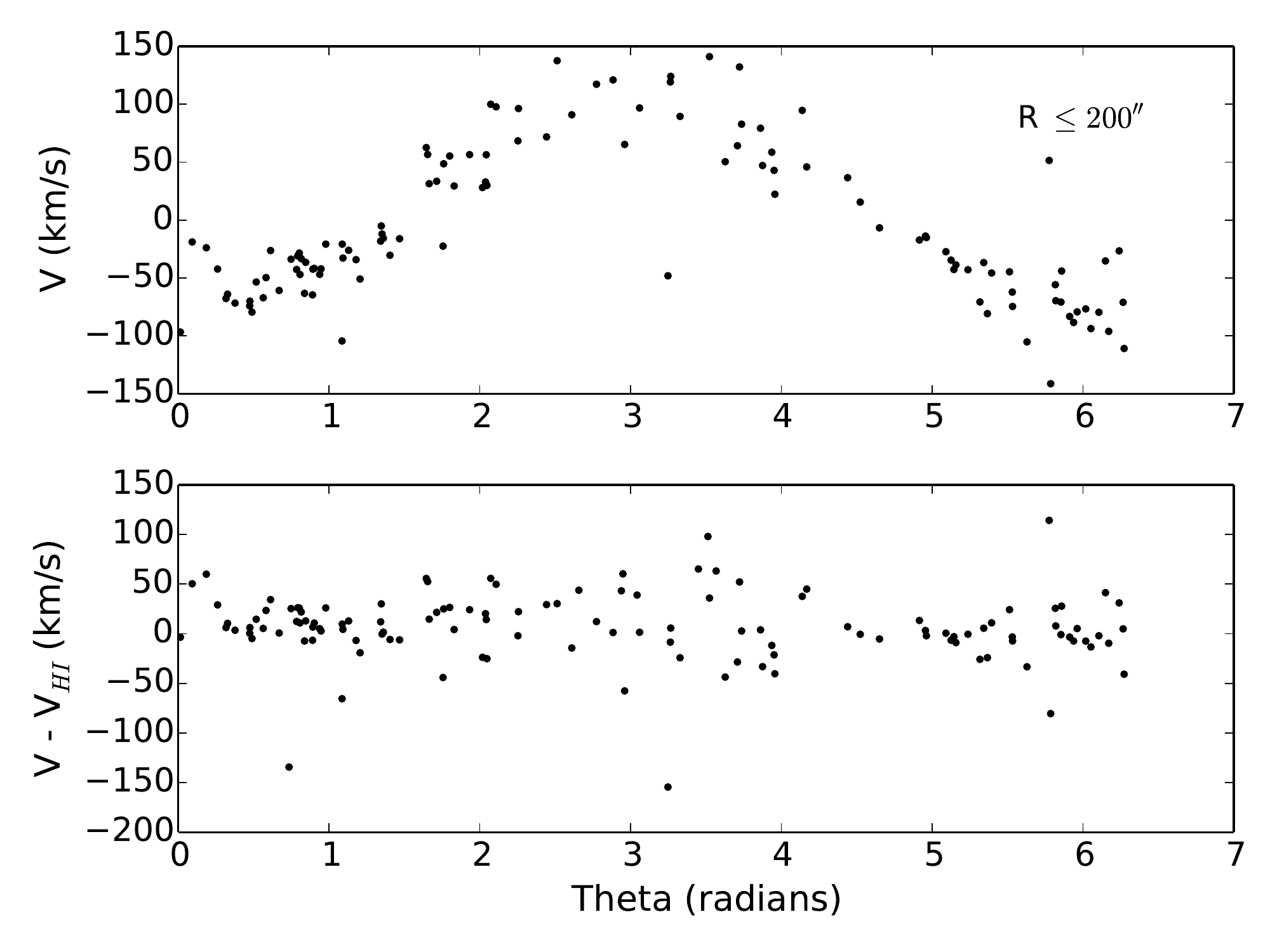}
   \caption{}
\end{subfigure}

\begin{subfigure}[b]{0.5\textwidth}
   \includegraphics[width=1\linewidth]{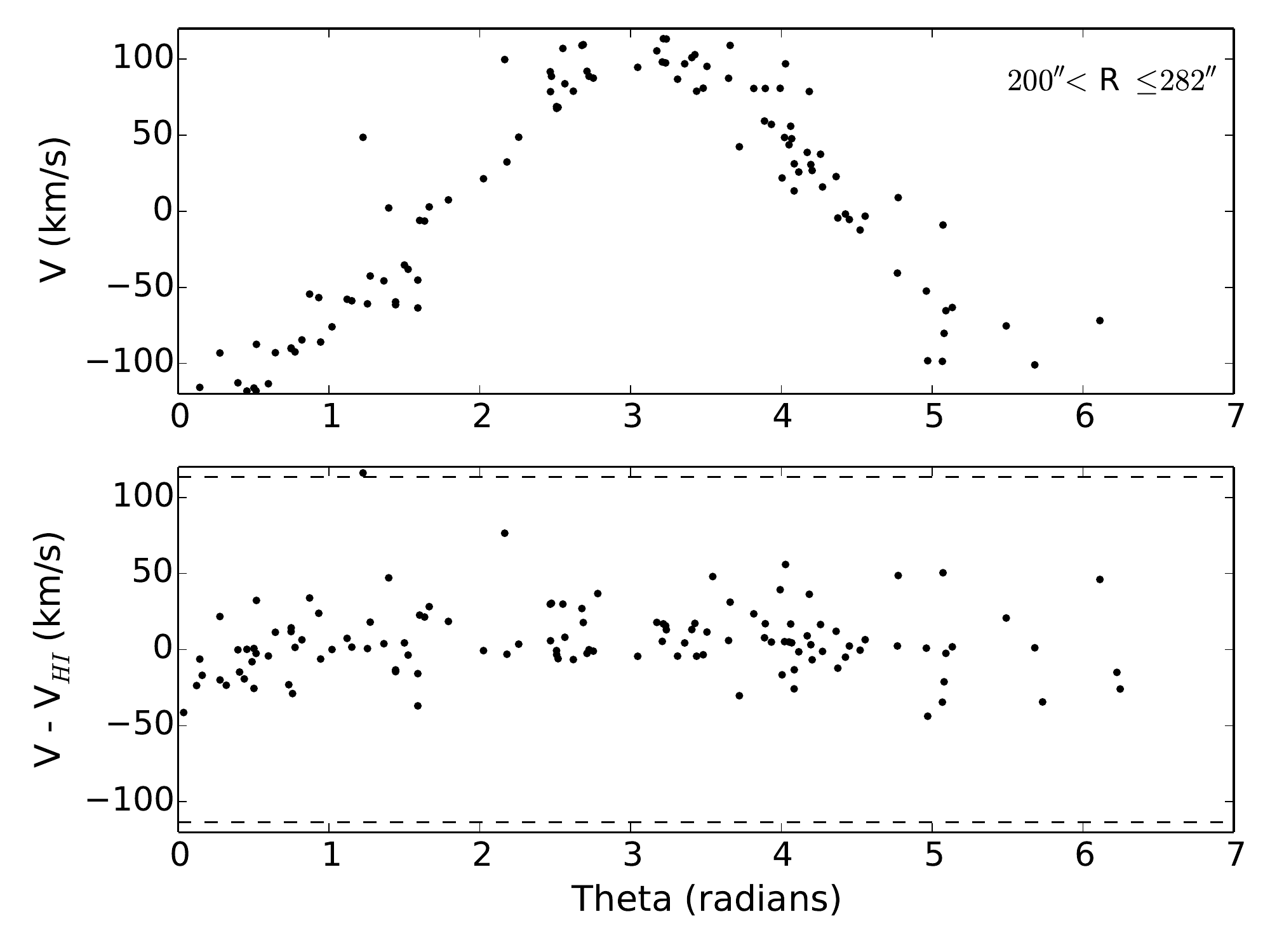}
   \caption{}
\end{subfigure}

\begin{subfigure}[b]{0.5\textwidth}
   \includegraphics[width=1\linewidth]{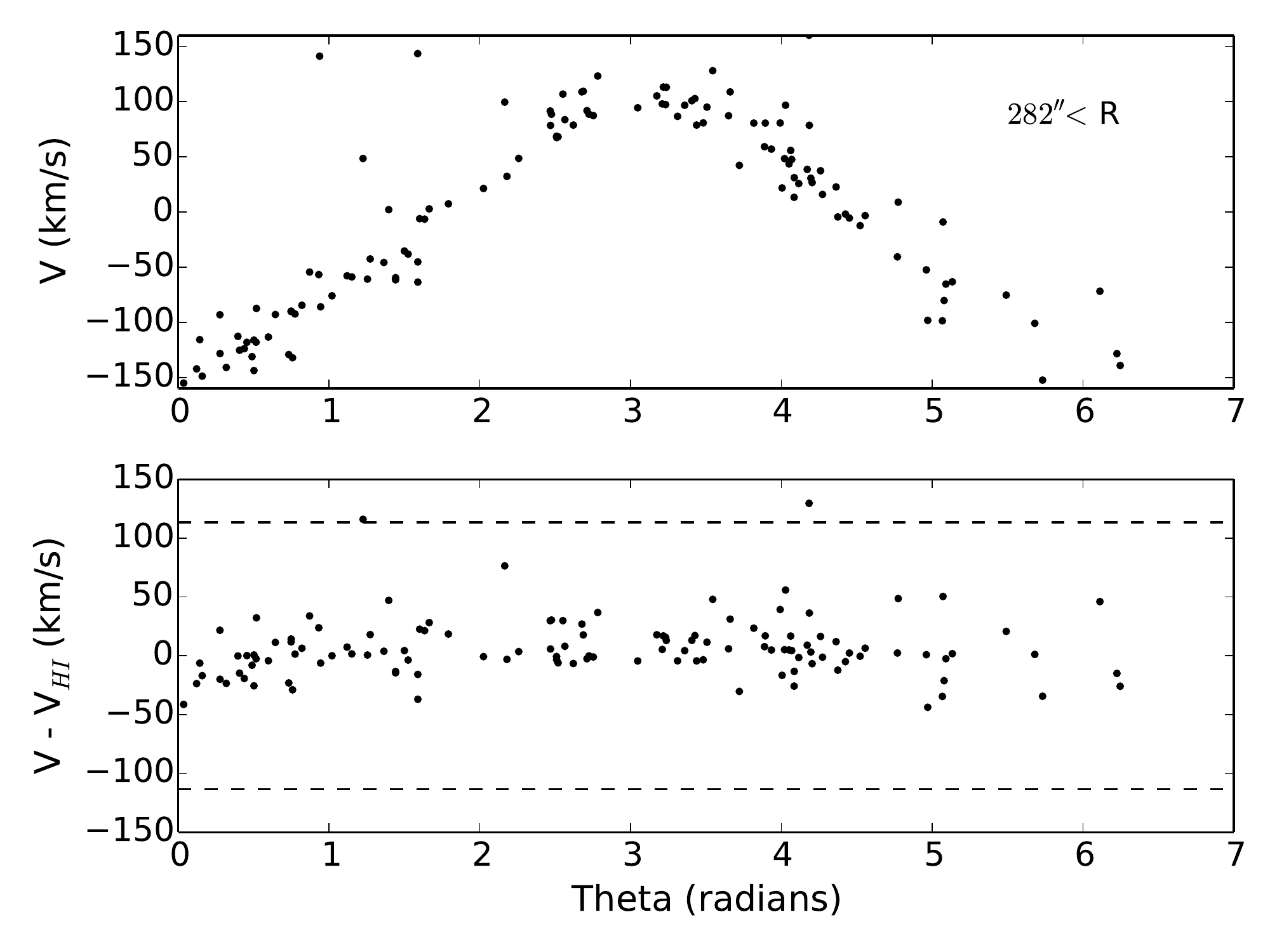}
   \caption{}
\end{subfigure}
\caption[Velocity vs azimuthal angle plots in 3 radial bins for PN.S data for NGC 6946.]{Velocity vs azimuthal angle plots in 3 radial bins, each with about 125 PNe. The angle $\theta$ is in the plane of the galaxy, measured from the line of nodes. The top panels show the velocity before taking off the local THINGS velocity and the bottom panels show the velocities after correcting for galactic rotation. A few outliers were removed based on a visual inspection of the velocity histogram of the objects in each bin. The objects within the dashed lines are the ones that were included in our sample for analysis; no objects were thrown out in the first bin. Each panel shows a mix of cold population of spatially unresolved emission line objects, plus a hotter population whose velocity dispersion appears to decrease with galactic radius.}
\label{vlos_theta_N6946}
\end{figure}

\begin{figure*} %fig 6
\includegraphics[width=1\linewidth]{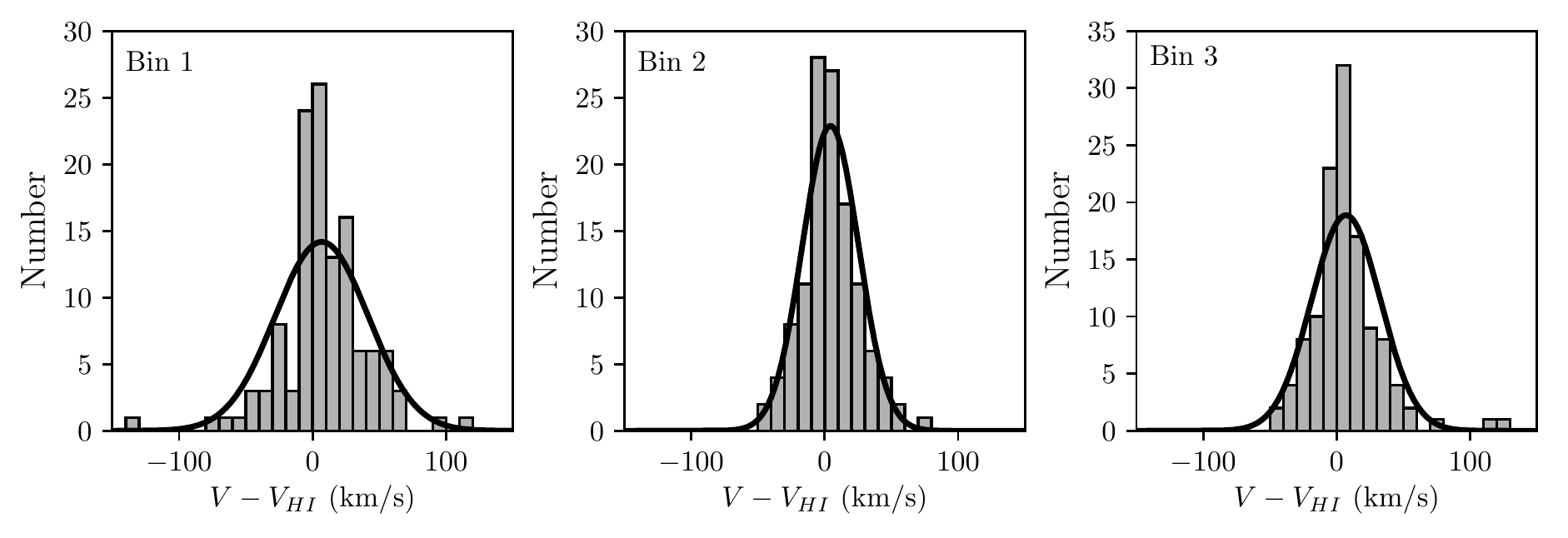}
\caption{The distribution of the PNe line-of-sight velocities from the bottom panels in Figure \ref{vlos_theta_N6946} for each radial bin respectively. The single Gaussian fit to the data is shown with the black curve. The velocity distributions are not well represented by a single Gaussian: the excess of low-velocity PNe is evident.}
\label{Hist_vlos}
\end{figure*}

\begin{table*} %table 3
\begin{tabular}{ccccccccccc}
\hline
Mean Radius  & \multicolumn{4}{c}{2 Component model}                       & \multicolumn{3}{c}{1 Component model} \\
            & $\sigma_{z,cold}$ & $\sigma_{z,hot}$ & BIC & $\chi^2_{red}$ & $\sigma_z$   & BIC  & $\chi^2_{red}$  \\
(arcsec)    & (\kms)            & (\kms)     &                   &     &           (\kms)               &     &                &       &      &                 \\
\hline
54          &        $26.6 \pm 4.9$           &     $65.3 \pm 12.1$             &        17164            &  1.225   &       $39.7 \pm 6.3$            &               17283 &       1.410                     \\
98          &     $19.3 \pm 3.6$              &       $64.6 \pm 12.2$           &      17351            &  1.231  &       $29.3 \pm  4.7$ &               17462 &       1.380                       \\
\hline
\end{tabular}
\caption[The single and double Gaussian fit from pPXF for NGC 6946.]{The single and double Gaussian fit from pPXF. For each component, the table gives the vertical velocity dispersion $\sigma_z$ for each of the components. Dispersions have been corrected for the contribution from the HI velocity dispersion. An estimate of the reduced $\chi ^2$ and the Bayesian Information Criterion parameter BIC is also given. Both the smaller chi-squared and the lower BIC suggest that the two component model is preferred.}
\label{VW_N6946}
\end{table*}

\begin{table*} %table 4
%\centering
%\resizebox{\columnwidth}{!}{
\begin{tabular}{cclccc}
\hline 
\multirow{2}{*}{\begin{tabular}[c]{@{}c@{}}Mean Radius\\ (arcsec)\end{tabular}} & \multicolumn{3}{c}{2 component Model} & \multicolumn{2}{c}{1 Component Model} \\
 & \begin{tabular}[c]{@{}c@{}}$\sigma_{z,cold}$ \\ (km s$^{-1}$)\end{tabular} & \multicolumn{1}{c}{\begin{tabular}[c]{@{}c@{}}$\sigma_{z,hot}$\\ (km s$^{-1}$)\end{tabular}} & BIC & \begin{tabular}[c]{@{}c@{}}$\sigma_z$\\ (km s$^{-1}$)\end{tabular} & BIC \\
\hline
144 & $\leq 12.1$ $\pm$ 2.3   &  32.0 $\pm$ 6.4  &1236  &  26.5 $\pm$ 1.7  & 1251  \\
242 &  $\leq 12.9$ $\pm$ 3.3  & 20.9 $\pm$ 4.1  & 1076  &  18.6 $\pm$ 1.2  &  1090\\
335 &  $\leq 12.3$  $\pm$ 3.5 & 14.8 $\pm$ 5.9   & 1062 &  14.5 $\pm$ 1.0  & 1055  \\
\hline
\end{tabular}
\caption[The $\sigma_z$ values calculated from the PN.S data for NGC 6946.]{The $\sigma_z$ values calculated from the PN.S data. We give the 90\% confidence upper limit for the cold dispersions. The lower BIC values of the two component fit suggest it to be the preferred model over the one component model, except for the outermost radial bin where the one component model is preferred.}
\label{PNS_N6946}
\end{table*}

\subsubsection{LOSVD and the Vertical Velocity Dispersion}
\label{sec_sas1}
We use pPXF \citep{Cappellari2017} to obtain the first two moments (the mean line-of-sight velocity and the $\sigma_{LOS}$) for the two Gaussians for the two VIRUS-W spectra. We fit two and single Gaussian components to the data for the comparison. The resulting fits are shown in Figure \ref{spec_N6946}.  Further, we use the Bayesian Information Criterion (BIC; \citealt{Schwarz:1978}) to judge the preferred model (one-component or two-component) for the spectra (see Eqn. 3 and Eqn. 4 in \citetalias{Aniyan:18} on how to calculate BIC). The BIC applies a penalty for models with a larger number of fitted parameters. Thus, between two models, the model with the lower BIC value is preferred. However, we note that in case of the similar BIC values, the BIC can only be considered as a suggested preference. For our data the two component fit is preferred over the single component fit in both of the radial bins, this result is also consistent with the reduced $\chi ^2$. The results for both fits are tabulated in Table \ref{VW_N6946}. %

pPXF uses a best-fit linear combination of stellar templates to directly fit the spectrum in pixel space and to recover the line of sight velocity distribution (LOSVD).  We observed template stars of different spectral types with VIRUS-W as our list of stellar templates, to avoid resolution mismatch between the stellar templates and the galaxy spectrum. We assume the two components of the LOSVD to be Gaussian for this close-to-face-on galaxy, and therefore retrieved only the first and second moment parameters from pPXF. 

pPXF finds an excellent fit to our spectrum, as shown in Figure \ref{spec_N6946}. It also returns the adopted spectra of the individual components, which are consistent with the spectra of red giants. The mean contributions of the cold and hot disc components to the total light are 52\% and 48\% in the inner radial bin and 46\% and 54\% in the outer bin. To correct $\sigma_{LOS}$ values for the contribution of the HI velocity dispersion we adopt a correction value to be $\sim$ 6 \kms\ %
(see Section \ref{sec_sas1a}).

We then calculate the vertical component of the stellar velocity dispersion $\sigma_z$ from the line of sight component $\sigma_{LOS}$ using the following equation:
\begin{equation}
\sigma_{LOS}^2 = \sigma_\theta^2 \cos^2 \theta . \sin^2 i + \sigma_R^2 \sin^2\theta .\sin^2 i + \sigma_z^2 \cos^2 i + \sigma_{meas}^2
\end{equation}
where $\sigma_R$, $\sigma_\theta$ and $\sigma_z$ are the three components of the dispersion in the radial, azimuthal and vertical direction, $\sigma_{meas}$ is the measurement error on the velocity and $i$ is the inclination of the galaxy ($i = 0$ is face-on).  %
Using the epicyclic approximation in the part of the rotation curve that is close to solid body, we adopt $\sigma_R = \sigma_\theta$,
% using the epicyclic approximation 
and for the part of the rotation curve that is flat, we adopt $\sigma_R$ = $\sqrt{2}\sigma_\theta$. Based on an examination of the THINGS HI velocities along the galaxy's kinematic major axis (last panel in Figure \ref{pa_N6946}) we determine this galaxy's rotation curve at radius $\leq 200''$ to be solid body and beyond $200''$ to have a flat rotation curve. We also adopt the stellar velocity ellipsoid parameter $\sigma_z/\sigma_R$ ratio to be $0.6 \pm 0.15$ following the result from \citet{Shapiro:03}. This value is consistent with the value used by \citet{Bershady:10b} and the value found in the solar neighbourhood by \citet{Aniyan16}. For the discussion regarding the uncertainties of $\sigma_z/\sigma_R$ measurements for external galaxies and its dependence on the morphological type please see \citetalias{Aniyan:18}, Section 5. The uncertainty in $\sigma_z/\sigma_R$ is included in the error of $\sigma_z$ as described in Section 7.1.2 of \citetalias{Aniyan:18}. We correct $\sigma_z$ values for the small broadening introduced by subtracting the local HI velocity to remove galactic rotation. Our results for the stellar $\sigma_z$ values for both stellar components are presented in Table \ref{VW_N6946}. The errors are the 1$\sigma$ errors obtained using Monte Carlo simulations. This was done by running 1000 iterations where, in each iteration random Gaussian noise appropriate to the observed SN of the IFU data was added to the best fit spectrum originally returned by pPXF. Then, pPXF was run again on the new spectrum produced in each iteration. The errors are the standard deviations of the distribution of values obtained over 1000 iterations.

\subsection{Planetary Nebulae}
\subsubsection{Removing Galactic Rotation}
From the PNe velocity field we also remove the effects of galactic rotation, similarly to the analysis of the IFU integrated light absorption spectra.
We use HI velocity at the position of %all 
each of our PNe from the THINGS first moment map and then subtract local HI velocities from the PNe velocities. 
These velocities, corrected for the galactic rotation, are henceforth denoted $\rm v_{LOS}$. We use $\rm v_{LOS}$ to calculate the velocity dispersions. As for the VIRUS-W data, the radius and azimuthal angle ($\theta$) of the PNe in the plane of the galaxy were calculated using our estimated PA and angle of inclination, and the $\rm v_{LOS}$ data were then radially binned into 3 bins, each with about 125 PNe. Figure \ref{vlos_theta_N6946} shows the $v_{LOS}$ vs $\theta$ plots in each radial bin before and after the HI velocities were subtracted off. The distribution of the $v_{LOS}$ after the correction for the HI velocities in shown Figure \ref{Hist_vlos} for 3 radial bins respectively. The distributions of the data points already suggest the presence of the two kinematically distinct components as it can not be well fit with the single Gaussian distribution, as shown in Figure \ref{Hist_vlos}, and show a clear indication of a cold kinematic component in $v_{LOS}$ in each radial bin.

\subsubsection{LOSVD and the Vertical Velocity Dispersion}  %\S 5.4.2.2
In each radial bin we remove a few obvious outliers, based on a visual inspection of the velocity histogram of the objects.  
 A maximum likelihood estimator (MLE) routine was then used to calculate the LOS velocity dispersions and the subsequent $\sigma_z$ in each radial bin. 
The first iteration in this routine estimates $\sigma_{LOS}$ for the kinematically cold and hot component by maximizing the likelihood for the two-component probability distribution function given by Eqn. 6 in \citetalias{Aniyan:18}.

In order to calculate the surface mass density using Eqn. \ref{main_eq2}, we need the vertical velocity dispersion of the hot component ($\sigma_{z}$). To determine it, we use a second MLE routine. Two parameters are passed to the function in this stage: $\sigma_{z1}$ and $\sigma_{z2}$ which are the vertical velocity dispersions of the cold and hot components respectively. The $\sigma_{LOS}$ values obtained using Eqn. 6 in \citetalias{Aniyan:18}  are passed to the routine as initial guesses, since the $\sigma_z$ will be very close to the value of $\sigma_{LOS}$ for this galaxy. We assume $f = \sigma_R / \sigma_z = 1.7\pm 0.42$ (see Section \ref{sec_sas1}) and use inclination $i = 37^\circ$ (see Section \ref{sec_rotcurve}). The PN.S data for the first radial bin are all at radii $< 200''$, where the rotation curve is close to solid body. In this bin, we assume the radial and azimuthal components of the velocity dispersion are equal i.e. $\sigma_R = \sigma_\theta$. The PN.S data for the second and third radial bin are all at radii $> 200''$ where the rotation curve is flat and we use: $\sigma_R = \sqrt{2}\sigma_\theta$, where $\sigma_R$ and $\sigma_\theta$ are the in-plane dispersions in the radial and azimuthal directions. 
Finally, once the initial guesses are passed to the routine, it calculates the expected $\sigma_{LOS}$ for the hot and cold component at each azimuthal angle ($\theta$) using the relation:
\begin{equation*}
\begin{split}
&\sigma_{LOS1}^2 = \frac{\sigma_{\text{z1}}^2f^2}{2} \cos^2\theta . \sin^2 i 
			    + \sigma_{\text{z1}}^2f^2 \sin^2\theta . \sin^2 i + \sigma_{\text{z1}}^2 \cos^2 i\\
&\sigma_{LOS2}^2 = \frac{\sigma_{\text{z2}}^2f^2}{2} \cos^2\theta . \sin^2 i 
			    + \sigma_{\text{z2}}^2f^2 \sin^2\theta . \sin^2 i + \sigma_{\text{z2}}^2 \text{cos}^2 i
\end{split}
\end{equation*}
The subscript 1 and 2 refers to the components of the cold and hot populations respectively. This step depends on the theta-distribution of the PN in each bin. Thus, the routine then proceeds to calculate the probability of a particular v$_{LOS}$ to be present at that azimuthal angle via the equation:
\begin{equation}
\begin{split}
P =  & \frac{N}{\sigma_{LOS1}\sqrt{2\pi}}\exp\left(\frac{-(\rm v_{LOS} - \mu_1 \cos \theta . \sin i)^2}{2\sigma_{LOS1}^2}\right ) + \\
& \frac{1-N}{\sigma_{LOS2}\sqrt{2\pi}}\exp\left (\frac{-(\rm v_{LOS} - \mu_2 \cos\theta . \sin i)^2}{2\sigma_{LOS2}^2}\right )
\end{split}
\label{sigdist}
\end{equation}
\begin{figure*} %fig 7
\centering
  \includegraphics[width=0.7\linewidth]{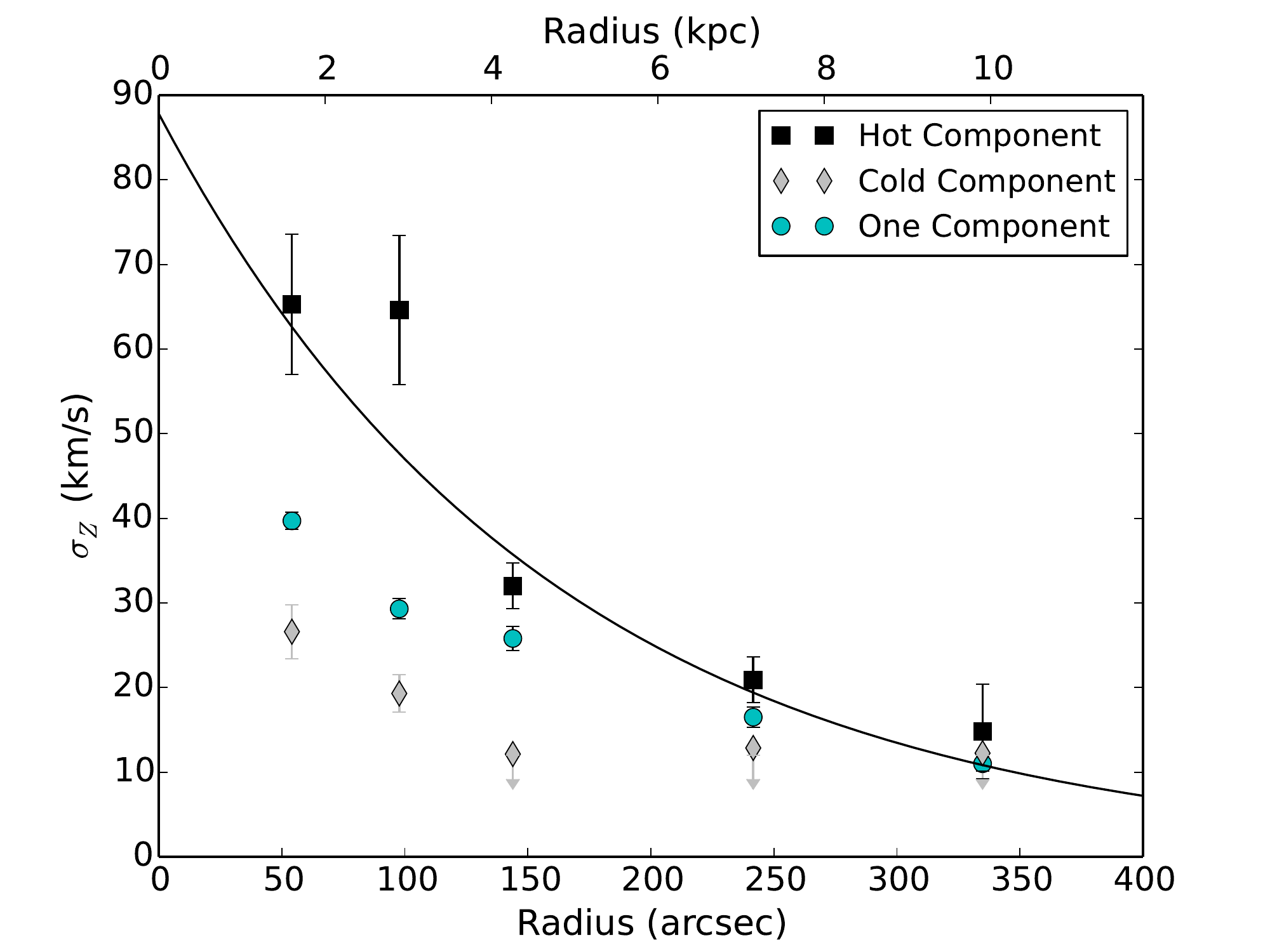}
\caption[The $\sigma_z$ per radial bin in NGC 6946.]{The vertical velocity dispersion as a function of radius in NGC 6946. The black and grey markers indicate the hot and cold velocity dispersions respectively from our two component fits, the cyan markers are our single-component values. The two inner data points at $R = 54''$ and $98''$ were obtained for integrated light spectra from VIRUS-W and the outer data points are for PNe from PN.S. The solid line denotes an exponential with twice the galaxy's dynamical scale length ($h_{dyn} = 92'' = 2.72$ kpc) fit to the hot component. Our data have been corrected for the HI velocity dispersion and PNe measuring errors. The cold component from the PN.S data were dominated by our errors, and we show the 90\% upper limits for these values. The errors bars are the 1$\sigma$ errors obtained from Monte Carlo simulations.}
\label{sigma_N6946}
\end{figure*}
In Eqn. \ref{sigdist} $N$ is the value of the fraction of the cold population returned by the routine that calculated $\sigma_{LOS}$ described earlier; $\mu_1$ and $\mu_2$ are associated with the means of the two components; the $\cos \theta . \sin i$ term factors for any asymmetric drift present in the data. Whether we let the code estimate the terms $\mu_1$ and $\mu_2$ or if we fix these terms at 0, it made negligible difference to the dispersions. This shows that there is negligible contribution from the asymmetric drift in the data in each of our radial annuli. Eqn. \ref{sigdist} is maximised to return the best fit values for $\sigma_z$ for the hot and cold population of PNe. The 1$\sigma$ errors are calculated similar to the method used in the analysis of the VIRUS-W data. We carry out our Monte Carlo error estimation by using the double Gaussian distribution found by our MLE code, to extract about 125 random velocities (i.e same number of objects as in each of our bins). We then use this new sample to calculate the $\sigma_z$ of the hot and cold component using our MLE routines. This whole process was repeated 1000 times, recording the dispersions returned in each iteration. The 1$\sigma$ error is then the standard deviation of the distribution of the dispersions returned from these 1000 iterations. The vertical velocity dispersions for the two components returned from the MLE routine is given in Table \ref{PNS_N6946}. After extracting $\sigma_z$ from $\sigma_{LOS}$, the presence of the cold component seems to remain prominent in bins 1 and 2, but not in bin 3, where the one component model is preferred. However, the distribution of the PNe line-of-sight velocities for bin 3 (Figure \ref{Hist_vlos}) still suggests the presence of the two kinematically distinct components.

The dispersions for the PNe were also corrected for the HI dispersion, similarly to the spectra analysis by quadratically subtracting HI velocity dispersion ($\sim$ 6 kms$^{-1}$) from the values of the vertical velocity dispersions returned by the MLE routine. This dispersion is evaluated by fitting a plane function to the HI velocities over the IFU and calculating the rms scatter of the HI velocities about this plane. We note that this scatter is not the same as the local HI velocity dispersion as measured from the HI profile width. We also correct the measured dispersions for the measurement errors of the individual PNe, as shown in Figure 3 in \citetalias{Aniyan:18}. The rms measuring error in each radial bin is about 6 km s$^{-1}$. The formal variance of the corrected cold dispersion value at all PNe radial bins is negative. Hence we show its 90\% confidence upper limit in Figure \ref{sigma_N6946} and Table \ref{PNS_N6946}. 
We carried out a comprehensive analysis using the colour-magnitude cut to discriminate between the  HII regions and PNe (Figure \ref{CMD_N6946}). Unfortunately, removing HII contaminants in disc galaxies is very challenging. There is a small probability that we may still have some HII contaminants in our cold component. However, we do not use the dispersion of our cold component in any further analysis. Our aim is to separate out the cold component (whether cold PNe or HII regions) from the kinematically hot component and then use the $\sigma_{z,\,hot}$ with the scale height for the same population to calculate the surface mass density. Hence, the nature of the objects that contribute to our cold dispersion is irrelevant as long as they are separated out effectively.

 \begin{table*} %table 6
 \begin{adjustbox}{width=2\columnwidth,center}
\begin{tabular}{cccccccccccc}
\hline
$R$ & $\sigma_z$ & $\Sigma_T$ & $\Sigma_{C,\,\rm gas}$ & $L_C/L_D$ & $F_C$ & $\Sigma_D$ & $\Sigma_{C,\,*}$ & $\Upsilon_{\star}$[B] & $\Upsilon_{\star}$[V] & $\Upsilon_{\star}$[I] & $\Upsilon_{\star}$[3.6]  \\ 
(kpc) & (\kms) & (M$_\odot$ pc$^{-2}$)  & (M$_\odot$ pc$^{-2}$)   &     &   & (M$_\odot$ pc$^{-2}$)  &(M$_\odot$ pc$^{-2}$)  & & & & \\ 
(1) & (2) & (3) & (4) & (5) & (6) & (7) & (8) & (9)& (10) & (11) & (12)  \\
\hline 

1.6  & 65.3 & 420$\pm$152   &   25.5$\pm$4   & 52/48 & 0.18 & 334.32  & 60.17 & 2.48$\pm$0.9 &  2.36$\pm$0.85  &  1.31$\pm$0.47   &    0.48$\pm$0.17\\  
2.9  & 64.6 & 411$\pm$157   &   17.7$\pm$4   & 46/54 & 0.14 & 345.0      & 48.3  & 3.61$\pm$1.2 &   3.61$\pm$1.2  &   2.07$\pm$0.69    &   0.72$\pm$0.24\\  
4.3  & 32.0 & 101$\pm$31      &  14.3$\pm$5  & 49/51 & 0.16 & 74.74    & 11.95 & 1.1$\pm$0.34  &  1.1$\pm$0.34     &    0.73$\pm$0.23   &   0.26$\pm$0.12\\  % 
7.2  & 20.9 &  43$\pm$16       &  8.3$\pm$4    & 49/51 & 0.16 & 29.91    &  4.78  & 0.91$\pm$0.34  &   1.12$\pm$0.41&    0.74$\pm$0.28   &   0.29$\pm$0.12\\  % 
9.9 & 14.8 &  21.5 $\pm$19    &  5.2$\pm$5    & 49/51 & 0.16 & 14.05      & 2.24 &   --      &    --     &   --      &   0.47$\pm$0.4\\  % 293"
\hline
\end{tabular}
\end{adjustbox}
\caption[Parameters for NGC 6946 at the five radii where the velocity dispersion of the hot disc was measured.]{ Parameters for NGC 6946 at the five radii where the velocity dispersion of the hot disc was measured  (inner two radii from integrated light, outer three from PNe). All surface densities are in units of \mspc. The columns are (1) radius (kpc); (2) vertical velocity dispersion $\sigma_z$ of the hot disc; (3) total surface density $\Sigma_T$ from Eqn. \ref{main_eq2}; (4) observed surface density $\Sigma_{C,\,\rm gas}$ of the gas layer; (5) ratio of luminosities of cold and hot layers from the integrated spectra in rows 1 and 2, and adopted mean of rows 1 and 2 in rows 3 to 5; (6) the ratio $F_C$ of the stellar surface densities of the cold and hot layers from \citet{Bruzual:03}; (7) the surface density $\Sigma_D$ of the hot layer from Eqns. \ref{app_gal11_N6946}; (8) the stellar surface density $\Sigma_{C,\,*}$ of the cold layer; (9)--(12) total ($\Sigma_D$+$\Sigma_{C,\,*}$) stellar mass-to-light ratio in BVI and 3.6$\mu$m bands. We note the absence of the last measured point of the $\Upsilon_{\star}$ in optical bands as our data do not extend far enough.}
\label{app_tab1_N6946}
\end{table*}

\section{Vertical velocity dispersion profile} % \S 7
\label{sec_dispersion}
Figure \ref{sigma_N6946} shows the % 
velocity dispersion results obtained from the integrated light VIRUS-W data (points at R = $54''$ and $98''$) and the planetary nebulae from the PN.S data (3 outer points) in each radial bin. At each radius we show %
the one component dispersion %
(cyan markers) and then the hot and cold thin disc dispersion from the double Gaussian fit (%
black and grey markers). 
The solid curve in Figure \ref{sigma_N6946} is an exponential fit to the hot component data in the form $\sigma_z\,(R) = \sigma_z\,(0)  \exp \,(-R/2h_{dyn})$, as expected from Eq. \ref{main_eq2} if the total surface density $\Sigma_T$ is exponential in radius and has a radially constant scale height $h_z$.  From this exponential fit we obtain the best-fit dynamical scale length to be $h_{dyn}$ = 92" which is in excellent agreement with the scale length measured for the 3.6 $\rm \mu$m band as $\rm I\,(R) =  I\,(0) \exp \,(-R/h_{r,[3.6]})$ (Table \ref{par_N6946}). 
The fitted central velocity dispersion for the hot component is $\sigma_z\,(0) = 87.8 \pm 4.9 $ \kms. We recall that it is this hot component dispersion that should be used with the derived scale height from Section \ref{sec_photometry} to estimate the total surface mass density of the disc. If we assume a single homogeneous population of tracers and fit the same exponential function to the one-component dispersions (cyan markers in Figure \ref{sigma_N6946}), we find the best-fit dynamical scale length to be $h_{dyn}$ = 120" and the central vertical dispersion $\sigma_z\,(0)=49.2 \pm 5.1$ \kms , which is $\sim 50 \%$ smaller than the central velocity dispersion from the fit to the hot component. The use of this one-component dispersion for the calculation of the surface mass density would underestimate the surface density by a factor of $\sim 2$, which would be enough to make the maximal disc look sub-maximal, but with a gradient in the mass-to-light ratio (see the results from \citet{HC3} about the increase in the disk mass-to-light ratio in the outer disc).

We note that the difference between the one component dispersions and the hot component dispersions decreases with radius and is almost zero in the outermost radial bin (Figure \ref{sigma_N6946}). This is consistent with the BIC values shown for the outer radial bin, which favours the one component model (Table \ref{PNS_N6946}).

\section{Stellar Surface Mass Density}
\label{sec:ssmd}
%\subsection{Isothermal model excluding the gravity of the cold component} %\S 5.3
Using Eqn. \ref{main_eq} with $f = 1/2\pi$ (isothermal model, see Appendix 1 in \citetalias{Aniyan:18}) and measured $\rm h_{z}$ (Table \ref{par_N6946}) and $\sigma_{z}$ from the hot disc (Table \ref{app_tab1_N6946}, column 2), we calculate the total surface mass density ($ \Sigma_T$) in each radial bin (Table \ref{app_tab1_N6946}, column 3). 

The stellar mass surface density of the hot disc ($\Sigma_D$) is $\Sigma_D = \Sigma_T - (\Sigma_{C,*} + \Sigma_{C,gas})$, where $\Sigma_{C,*} + \Sigma_{C,gas}$ is the surface density of a cold thin layer made up of the cold thin disc of stars and the gaseous disc of the galaxy. While we can directly measure $\Sigma_{C,gas}$ (see Figure \ref{cold_gas} and Table \ref{app_tab1_N6946}, column 4), the cold stellar population contribution $\Sigma_{C,\,*}$ is not known directly from our data. 
We can write the equation for the surface densities of the stellar disc components as:
\begin{equation} 
\Sigma_D = \frac{\Sigma_T - \Sigma_{C,\,\rm gas}}{1 + F_C},~~{\rm where}~~~ F_C\,= \frac{\Sigma_{C,\,*}}{\Sigma_D}. 
\label{app_gal11_N6946}
\end{equation}  

We can estimate the ratio of luminosities $L_C$/$L_D$ of the cold and hot stellar layers from the integrated spectra. These luminosity ratios are given in Table \ref{app_tab1_N6946} (column 5) for the first two radial positions where the dispersions come from integrated light. From these ratios, we can then estimate the ratios of the surface densities of the cold and hot layers, using stellar population models \citet{Bruzual:03}.   

\begin{figure} %fig 5
\includegraphics[width=1.1\linewidth]{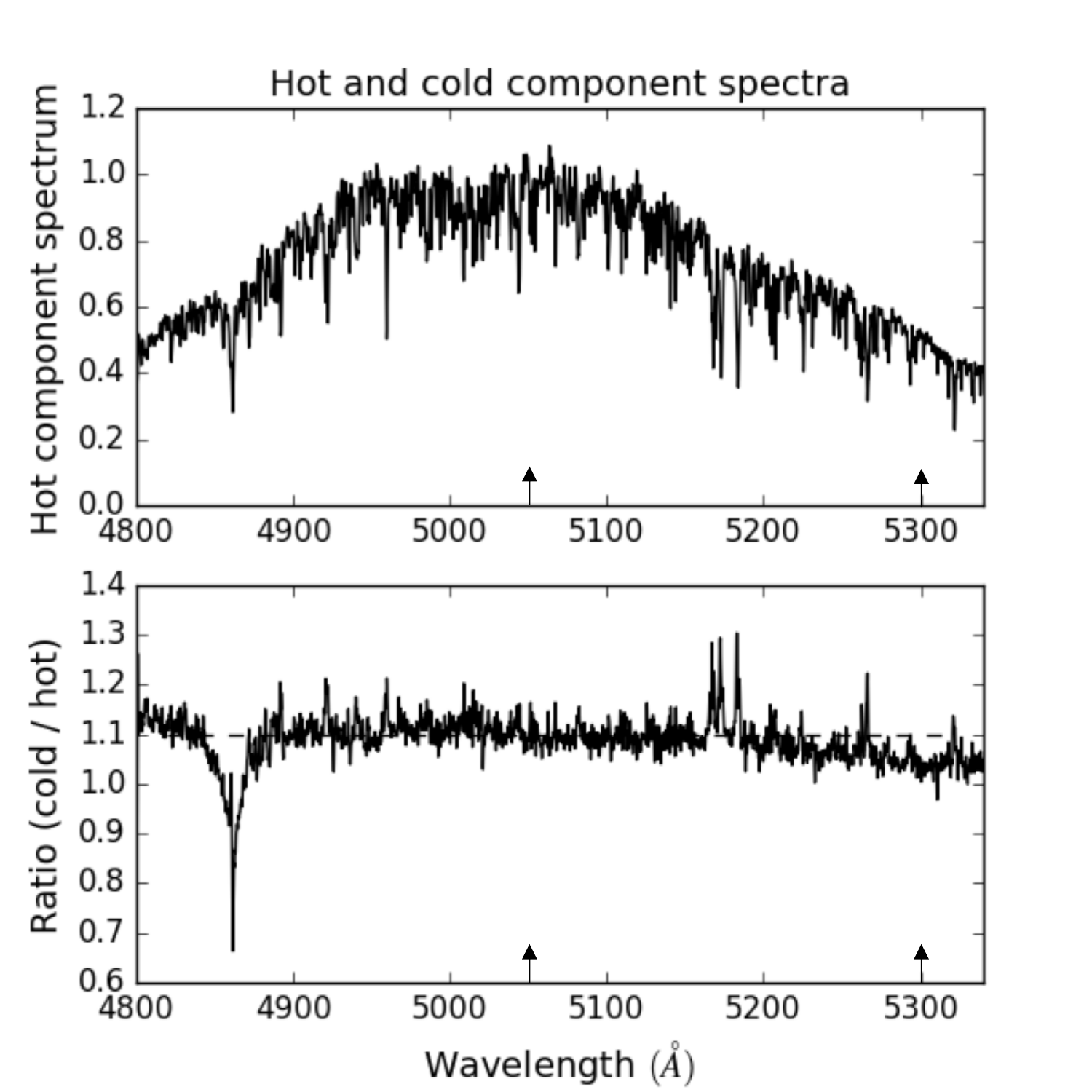}
\caption{The upper panel shows the spectrum for the hot component for the inner radial zone in NGC 6946. The lower panel shows the ratio of the cold to the hot spectra, pixel by pixel. The spectral region between 5050 and 5300 \AA \ used for the fit is indicated with vertical arrows. The [NI] emission lines at $\sim$ 5200 \AA \ have been omitted from the fit.}
\label{hot_cold}
\end{figure}

We showed in \citetalias{Aniyan:18} (Figure 12) that the thin disc stars in the solar neighbourhood older than about 3 Gyr have vertical velocity dispersions almost independent of age, at about $20$ \kms. Stars with ages younger than about 3 Gyr show a strong age-velocity dispersion relation, rising from about $10$ \kms\ for the youngest stars to about $20$ \kms\ at 3 Gyr. We identify stars with ages $> 3$ Gyr as the old hot population denoted D in Table \ref{app_tab1_N6946}, and stars with ages $< 3$ Gyr as the young cold populations denoted C$_*$ in Table \ref{app_tab1_N6946}. 

The age difference of the cold and hot stellar populations is also indicated visually by the spectra of the best-fit stellar templates used in the spectral fitting analysis (Figure \ref{spec_N6946}): the colder population looks younger, with its stronger $H_\beta$ and (slightly) weaker metallic lines. Figure \ref{hot_cold} shows an extended section of the pPXF spectra including the H$\beta$ region. This region  was excluded from the kinematical analysis, thus is not shown in Figure \ref{spec_N6946}. Three trends are evident in the ratio of the cold/hot spectra: a) H$\beta$ is stronger in the spectrum for the kinematically colder component, and shows broader wings;  b) the metallic lines such as the Mgb triplet and the 5270 \AA\ and 5320 \AA\ Fe-lines are stronger in the kinematically hotter component and c) the continuum for the colder component is somewhat bluer than for the hotter component, as seen by reference to the horizontal dashed line. These three trends are consistent with the colder population being younger, and follow naturally if the mean temperatures of the stars which dominate the spectrum are lower in the hotter component. We note that only the region of spectrum from 5040 to 5310 $\rm \mathring{A}$ was used in the pPXF fit.

In \citetalias{Aniyan:18}, we then showed that, if the star formation rate in the disc has decayed with time like $\exp\,(-t/\tau)$, and if $\tau > 3$ Gyr, then the ratio of  ($M/L$ for the cold component) to ($M/L$ for the hot component) is about $0.167$, almost independent of $\tau$.
Assuming NGC~6946 has a similar age-velocity relation as in the solar neighbourhood, i.e the old thin disc is composed of stars with ages between 3 -- 10 Gyr and assuming a star formation rate that decays with time like $\exp(-t/\tau$), then we can derive the ratio of surface densities of young and old populations given as $F_C$ in Table \ref{app_tab1_N6946} (column 6) for the first two radial positions.  For the three radial positions using planetary nebulae as tracers ($R = 4.3$ to $9.9$ kpc), we cannot use these arguments, because of possible contamination of the sample by HII regions, and because the lifetimes of planetary nebulae vary strongly with progenitor mass \citep{miller16}. For columns 5 and 6 of Table \ref{app_tab1_N6946}, we therefore adopt the means of the values found from our integrated light spectra in the first two radial positions. The values of $\Sigma_D$ and $\Sigma_{C,*}$ in columns 7 and 8 follow from Eqn. \ref{app_gal11_N6946}. The errors for $\Sigma_T$, shown in Table \ref{app_tab1_N6946} are relatively large (30 to 40 \% for most radial bins) - these errors include errors in all spectroscopic and photometric parameters (scale length and scale height) used to evaluate $\Sigma_T$, and therefore following simple error propagation overestimate the relative errors between radial bins.

Table \ref{app_tab1_N6946} (columns 9 to 12), gives the total stellar mass-to-light ratios for ($\Sigma_D + \Sigma_{C,*})$ at each radius in BVI and 3.6 $\mu$m photometric bands (Figure \ref{fig_SBPS}). Prior to derivation of the mass-to-light ratio all photometric magnitudes were corrected for the Galactic extinction and inclination effects.
From Table \ref{app_tab1_N6946} it is visible that values of the mass-to-light ratio (M/L, $\Upsilon_{\star}$) vary with radius in every band differently from some previous studies \citep{martinsson13,swaters14}. In our approach, even if the light declines in the same way as mass ($\rm h_{phot}=h_{dyn}$), the ratio of the cold-to-hot component and the contribution from gas mass also changes with radius, contributing to radial variation of the M/L values. However, we also do not exclude the contribution of the observational errors to this radial change in M/L. Interestingly, the mean value of the $\Upsilon_{\star}$[3.6] is equal to 0.4, which is lower in comparison with the current stellar population models which assume constant $\Upsilon_{\star}$[3.6]=0.6 \citep{meidt12,rock15}. This lower value of $\Upsilon_{\star}$[3.6] can be explained with the very high star formation rate for this galaxy ($4.76 \, M_\odot yr^{-1}$, \citealp{things}). As was shown by \citet{miguel15} the flux of the Spitzer 3.6 $\mu$m band represents not only the light from the old stellar population, but also that emitted by the warm dust heated by young stars and re-emitted at longer wavelengths, and its contribution can be as high as 30\% at the regions of high star formation activity. Moreover AGB stars also peak at 3.6 $\mu$m, significantly contributing into the the total flux, but not into the stellar mass of a galaxy. \citet{ponom17} also shown that the use of the 3.6 $\mu$m luminosities corrected for the non-stellar contamination can even decrease the scatter in the Tully-Fisher relation. Thus, if we assume the non-stellar contamination of $\sim$ 30 \% for NGC~6946, the values of the $\Upsilon_{\star}$[3.6] will increase towards the mean value of $\sim$ 0.7. Moreover, the mean value of $\Upsilon_{\star}$[3.6]=0.4 is in agreement with the $\Upsilon_{\star}$[3.6] derived as a function of the [3.6]--[4.5] colour for a sample of spiral galaxies (\citealp{ponomareva18}, see Eqn.13.)  
In comparison with NGC~628 \citepalias{Aniyan:18} we find the mean $\Upsilon_{\star}$ to be lower for the NGC~6946 independently of a photometrical band. This is consistent with the difference in total star formation rate between the two galaxies: \citet{things} give the SFR for NGC 628 as  $1.21 M_\odot \,yr^{-1}$  and for NGC 6946 as $4.76 M_\odot \,yr^{-1}$, indicating that the stellar population in NGC 628 is likely to be older in the mean. This is also consistent with the above-mentioned non-stellar contamination in the 3.6 $\mu$m band. 

\section{Rotation Curve Decomposition}  %\S 9
\subsection{Observed HI rotation curve}
\label{sec_rotcurve}
\begin{figure} %fig 4
\centering
\includegraphics[width=\linewidth]{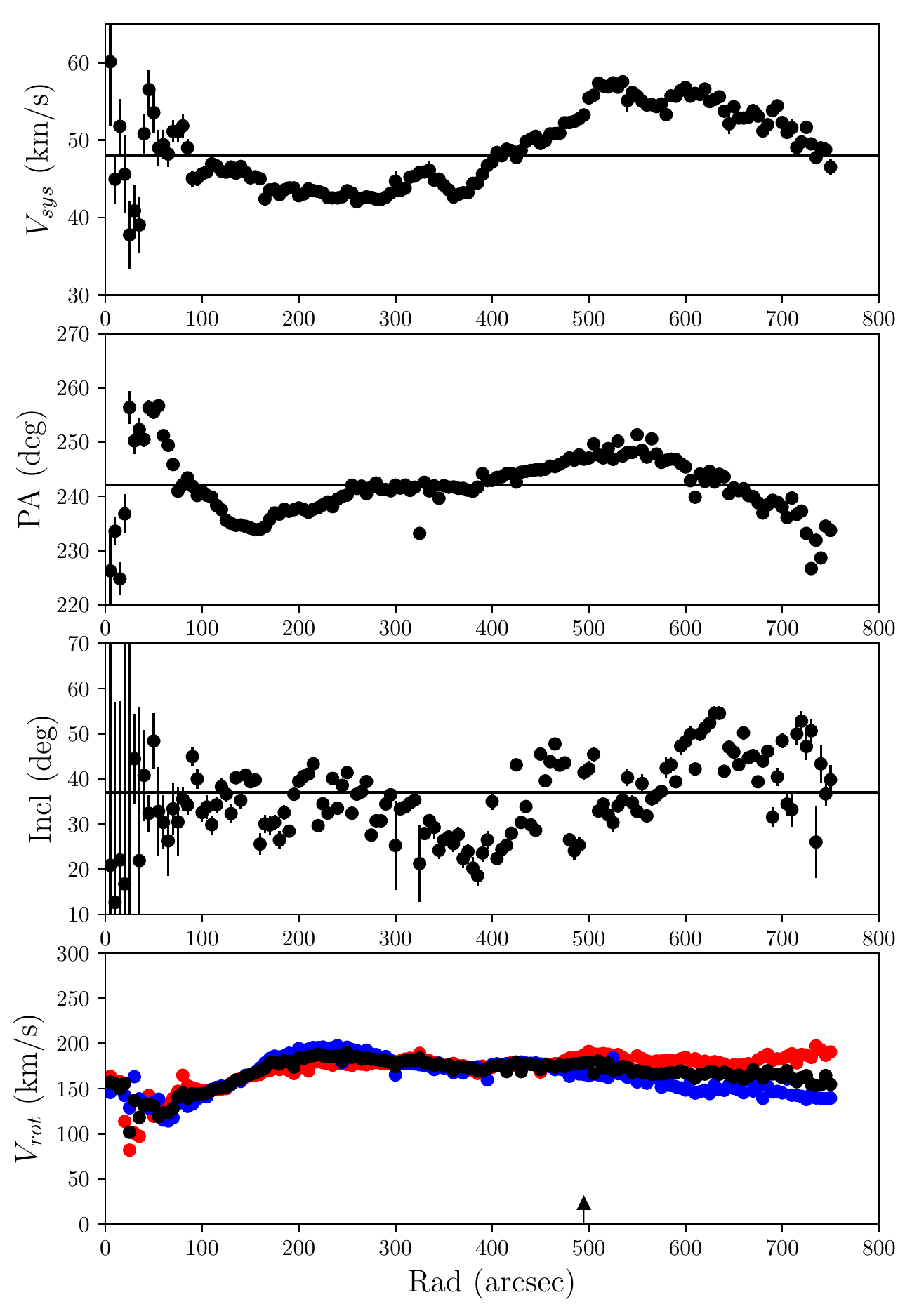}
\caption[Kinematic fits to HI THINGS data for NGC 6946]{Kinematic fits to the HI THINGS data for NGC 6946. The top three panels show the systemic velocity, position angle and inclination as a function of radius. The black solid lines indicate the average values which were used in the fit for the rotational velocity. The bottom panel shows the HI rotation curve of the galaxy. The blue and red data points are for the approaching and receding side of the galaxy respectively. The black curve is the average rotation curve for both sides. The optical radius (R$\rm_{opt} \sim 3.2 \rm h_{R}$) of the galaxy is indicated with the black arrow.}
\label{pa_N6946}
\end{figure}

We derive the observed rotation curve of NGC~6946 using THINGS data \citep{things}. As the galaxy is large and well-resolved we use the 2D tilted-ring modelling approach \citep{begeman89} to derive the rotation curve from the observed velocity field. First the velocity field was constructed using % 
the Gauss-Hermit polynomial fitting function and then corrected for the skewed velocity profiles which reflect random motions \citep{ponomareva16}. The results of the tilted-ring modelling are shown in Figure \ref{pa_N6946}. We find the position and inclination angles to be 242 and 37 degrees respectively. We fix these values to derive the final rotation curve for the receding and approaching sides of the galaxy, shown in the bottom panel of Figure \ref{pa_N6946}. It is clear that the rotation curve of NGC~6946 is well-behaved, reaching the flat part at $\sim$ 170". The difference between the approaching and receding sides of the rotation curve was adopted as the error on the rotational velocity. 

\begin{figure*} %fig 9
\centering
   \begin{subfigure}[b]{\textwidth}
   \includegraphics[width=\linewidth]{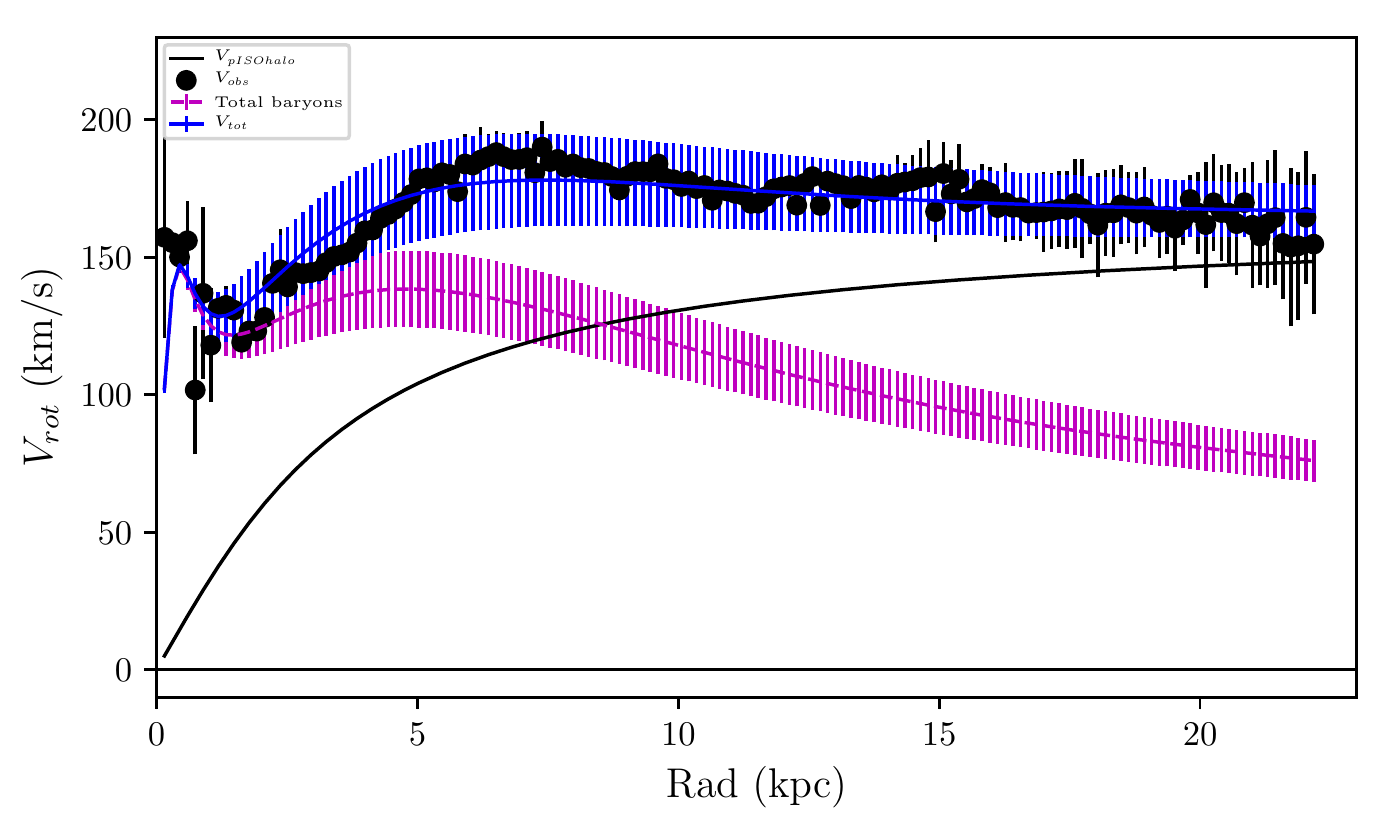}
\end{subfigure}

\begin{subfigure}[b]{\textwidth}
   \includegraphics[width=\linewidth]{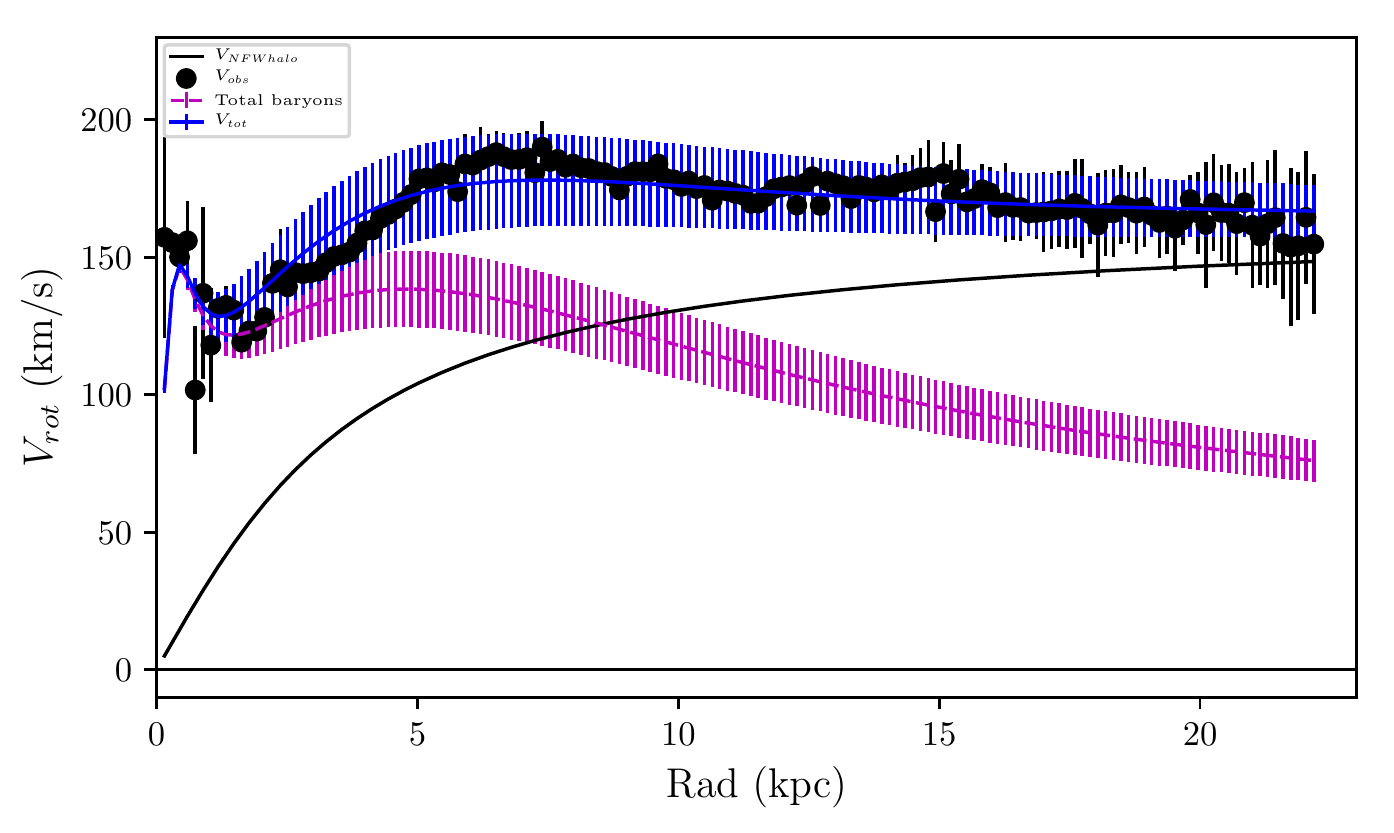}
\end{subfigure}

\caption[The rotation curve decomposition for NGC 6946.]{The rotation curve decomposition for NGC 6946 by fitting a pISO halo (top panel) and a NFW halo (bottom panel).
The observed HI rotation curve is shown as black dots. The magenta dashed line is the rotation curve from the total baryonic component: bulge+gas+disc with the $1\sigma$ errors. The blue line represents the total rotational velocity: $V^{2}_{tot}=V^{2}_{bar}+V^{2}_{halo}$.
The short blue vertical lines show the $1\sigma$ errors of the modelled total rotational velocity. $\rm 1kpc=33.8''$.}
\label{fig_halos_N6946}
\end{figure*}

\subsection{Stellar distribution}
The rotation curve and the 3.6$\mu$m surface brightness profile of NGC~6946 indicate the presence of the small bulge component. As our spectral observations do not cover the central region of the galaxy, the bulge does not contribute to the derived total surface mass density. Therefore, to include %
the contribution from the bulge % in 
to the total observed rotation curve, we fit the bulge component to the 3.6 $\mu$m profile (Figure \ref{fig_SBPS}) and then convert its luminosity into mass using $\Upsilon_{\star}[3.6]=0.6$ \citep{meidt12, miguel15,rock15}. 

Thus, we have all of the baryonic components contributions and we model their rotation curves using % 
a spherical potential for the bulge and the exponential total disc component, using % obtained 
the derived central surface mass density ($\Sigma_{0}=758.84 \pm 162$ M$_\odot$ pc$^{-2}$), dynamical scale length ($\rm h_{dyn}=2.72$ kpc) and I-band scale height ($\rm h_{z}=376$ pc), inferred from the measured I-band scale length (Section \ref{sec_photometry}). We model the rotation curves of all the baryonic components with the same radial sampling as was used for the derivation of the observed rotation curve. 

\subsection{Mass modelling} %\S 9.3
The total rotational velocity of a spiral galaxy can be presented as the %
quadratic sum of %
the rotational velocity for its baryonic components and a dark halo:
\begin{equation}
V^{2}_{tot}=V^{2}_{bar}+V^{2}_{halo},
\end{equation}

where $V_{bar}$ is the rotational velocity of the baryonic components of a galaxy (stars and gas) and $V_{halo}$ is the velocity of a dark matter halo.
From our analysis we have direct measurements on all baryonic components of the disc, using dynamically obtained surface mass density, and of the bulge from the 3.6$\mu$m profile. Thus we can model the total baryonic rotational velocity curve and then fit the rotation curve of the dark matter halo %until 
so that $V_{tot}$ matches $V_{obs}$ as closely as possible.
Since we have directly measured total surface mass density of the disc we do not have % any of the 
the usual uncertainties related to the disc $M/L$ that are a major concern in decomposing rotation curves.

We estimate the maximum rotational velocity of the baryonic rotation curve (blue line in Figure \ref{fig_halos_N6946}) at $\rm 2.2h_{dyn}$ to be equal to $\rm V_{max}(bar)=130$ km s$^{-1}$ with the typical error of 13\% due to the error on the central surface brightness and the negative covariance between the $ \rm h_{R,dyn}$ and the $\sigma_{z}(0)$ (see \citetalias{Aniyan:18} for more details). In comparison with the maximum velocity of the total rotation curve at $ \rm 2.2h_{dyn}$ ($ \rm V_{max}=170\pm 10$ km s$^{-1}$) we find our disc to be closer to maximal with $\rm V_{max}(bar)=0.76(\pm0.14)V_{max}$.

\begin{figure} %fig 11
\centering
   \begin{subfigure}[b]{\textwidth}
   \includegraphics[width=0.5\linewidth]{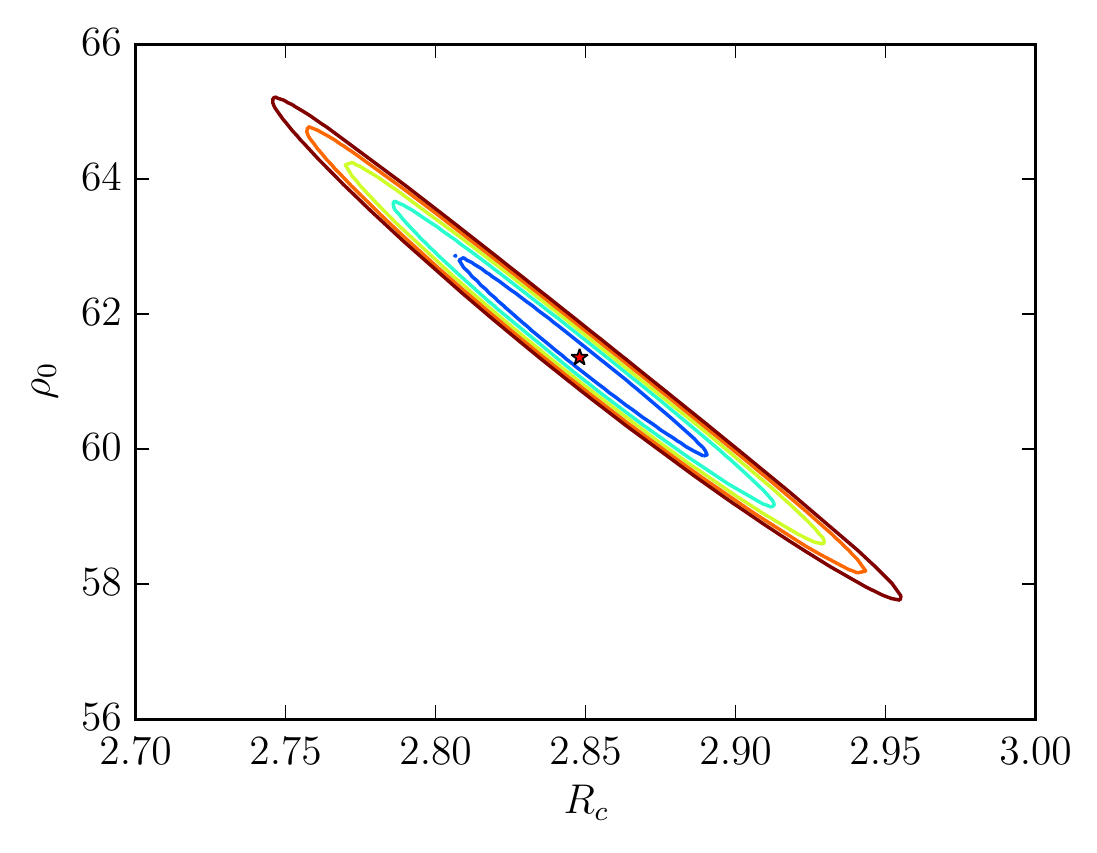}
\end{subfigure}

\begin{subfigure}[b]{\textwidth}
   \includegraphics[width=0.5\linewidth]{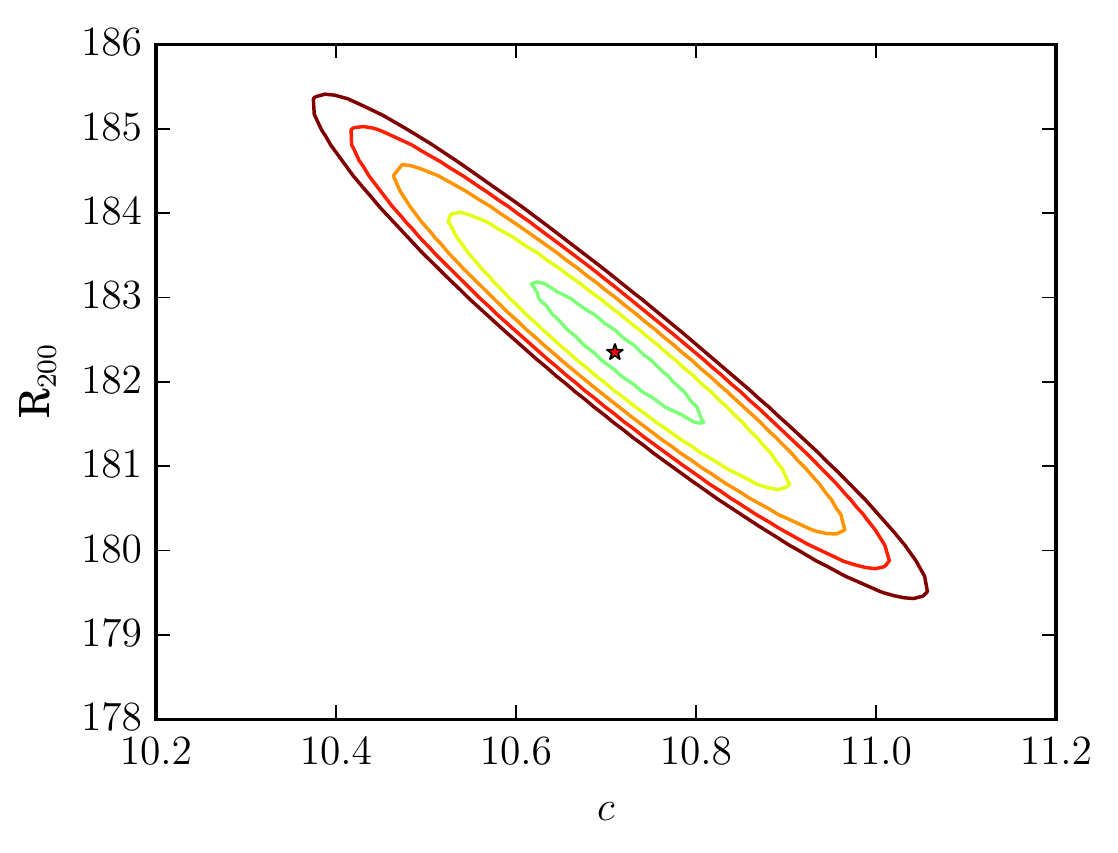}
\end{subfigure}
\caption{$\chi^2$ map in the 2D parameter space of the dark halo parameters. The upper panel shows the covariance between the density and radius of the core of the pseudo isothermal model of the dark halo. The lower panel shows the covariance between the virial radius and concentration of the NFW dark halo model. The ellipses represent the 1$\sigma$-5$\sigma$ values from inside out.}
\label{fig_chi2}
\end{figure}

For our further analysis we use two different % types of the 
dark matter halo models: the pseudo isothermal (pISO) and the NFW halo. 
The pISO halo rotation curve is parametrised by its central core density $\rho_{0}$ and % the radius of the core 
its core radius $R_{c}$:
 \begin{equation}
V_{DM}^{pISO} (R)=\sqrt{4\pi G \rho_{0}R^{2}_c \Big[1-\frac{R_c}{R} {\rm tan}^{-1}\Big(\frac{R}{R_c}\Big)\Big]},
\end{equation}
while the NFW halo \citep{navarro97} is parameterised by its %mass ($M_{200}$) 
circular velocity $V_{200}$ at the % within the 
virial radius $R_{200}$ and its central concentration $c$:
\begin{equation}
V_{DM}^{NFW} (R)=V_{200}\bigg[\frac{ln(1+cx)-cx/(1+cx)}{x[ln(1+c)-c/(1+c)]}\bigg]^{1/2}
\end{equation}
where $x=R/R_{200}$. % and $V_{200}$ is the circular velocity at $R_{200}$. 

Thus, we keep the rotation curve of the baryons fixed, and use %
the Gipsy \citep{gipsy} task ROTMAS to fit the rotation curves for the dark haloes. The results from this modelling is presented in Figure \ref{fig_halos_N6946} (the top panel is for the pISO halo and the bottom panel for the NFW halo). The derived parameters of the fitted dark matter rotation curves are presented in Table \ref{tbl_halos_N6946}. The error on the total rotational velocity (blue line in Figure \ref{fig_halos_N6946}) follows from the error on the central surface density.
Although the rotation curve shapes for the pISO and NFW halos are similar, the pISO % has a preference for our fitting model 
rotation curve is a slightly better fit, with $\chi_{red}^{2}=0.93$ versus $\chi_{red}^{2}=1.1$ for the NFW halo. 

Figure \ref{fig_chi2} shows the $\chi^2$ maps of the 2D parameter space between the main parameters of the dark matter haloes. It is clear that derived dark halo parameters for the pseudo isothermal and NFW models have significant covariance: the coloured ellipses in each panel represent the 1$\sigma$-5$\sigma$  values from inside to outside. The errors quoted in Table \ref{tbl_halos_N6946} are the 1$\sigma$ uncertainties.

\begin{table} %table 7
%\centering
\begin{tabular}{cccccc}
\hline
\multicolumn{3}{c}{pISO}&\multicolumn{3}{c}{NFW}\\
$R_{C}$&$\rho$&$\chi_{red}^{2}$&C&$R_{200}$&$\chi_{red}^{2}$\\
(kpc) & ($10^{-3}$ M$_\odot$pc$^{-3}$) & & &(kpc) & \\
\hline
 2.8 $\pm$ 0.2&57.6 $\pm$ 8&0.93&10.8 $\pm$ 1&169.3 $\pm$ 7&1.1\\
\hline
\end{tabular}
\caption[Derived dark halo parameters from mass modelling in NGC 6946.]{The derived parameters of the fitted dark matter haloes from the mass modelling. Solely based on the value of the $\chi^2_{red}$, the pISO model may be a better fit to the data.
}
\label{tbl_halos_N6946}
\end{table}

\section{Conclusions} %\S 6
We use absorption line spectra in the inner regions and planetary nebulae in the outer regions of the galaxy NGC 6946 to trace the kinematics of the disc. We show that there exists a younger, kinematically colder population of tracers 
%among a more
within an older and hotter component.

When attempting to break the disc-halo degeneracy by measuring the surface mass density of the disc, using the velocity dispersion and the estimated scale height of the disc, it is crucial that the dispersion and the scale height % refer 
pertain to the same population of stars. The scale height, obtained from NIR studies of edge-on galaxies % pertains to 
is for the older population of thin disc stars. We use this scale height with the dispersion of the hotter component to calculate the surface mass density. This density is a factor of 2.3 times greater than the surface density we would get if we assume a single homogeneous population of tracers. This factor is large enough to make the difference between concluding that a disc is maximal or sub-maximal. 

We find that the observed vertical velocity dispersion of the hotter component follows an exponential radial decrease. In comparison, the central velocity dispersions for the hot components in NGC 628 and NGC 6946 are not differ by much: 73.6 $\pm$ 9.8 km s$^{-1}$ for NGC 628 and 87.8 $\pm$ 4.9 for NGC 6946. The B-band magnitudes from \citet{things} are -19.97 and -20.61 respectively, suggest that the brighter galaxy has a higher central $\sigma_z$.

 The dynamical scale length of the galaxy (derived by fitting $\sigma_z\,(R) = \sigma_z\,(0)  \exp \,(-R/2h_{dyn})$ to the hot component velocity 
dispersions in Figure \ref{sigma_N6946}) agrees well with the photometric scale length of the galaxy (derived by fitting $\rm I\,(R) =  I\,(0) \exp \,(-R/h_{r,[3.6]})$  to the 3.6 $\mu$m surface brightness distribution in Figure \ref{fig_SBPS}). In this case the expected $M/L$ in this galaxy should be close to constant over the radial region probed by our study (see Section \ref{sec:ssmd}). We find that in all four photometric bands (BVI and 3.6 $\mu$m) the $M/L$ varies with radius, but within the errors being consistent with a radially constant value. Moreover, we find that $M/L$  in the 3.6 $\mu$m band  has a lower value than assumed by single stellar population models \citep{meidt12, miguel15,rock15}. Interestingly, its mean value over all radii agrees well with the $\Upsilon_{\star}$[3.6] derived as a function of the [3.6]--[4.5] colour  for a sample of spiral galaxies \citep{ponomareva18}. We suggest that lower values of the $\Upsilon_{\star}$[3.6] are due to the contamination from dust and AGB stars of the 3.6 $\mu$m flux. The maximum correction of 30 \% \citep{miguel15} would increase the mean $\Upsilon_{\star}$[3.6] by a factor of two. 

Decomposing the rotation curve of this galaxy, after taking into account the hot and cold stellar components, leads to a maximal disc ($\rm V_{max}(bar)=0.76(\pm 0.14)V_{max}$). The disc contributes about 76\% of the rotation curve at its peak, which is consistent with our previous study of NGC 628 (78\%). The molecular gas makes an unusually large contribution in the inner parts of this galaxy, and the baryons together dominate the radial component of the gravitational field out to a radius of about 8 kpc.
 
\section*{acknowledgements}
We thank Maximilian Fabricius for his help with the VIRUS-W
data reduction and analysis. We thank the 
anonymous referee for comments and suggestions which significantly 
improved this paper. We acknowledge comments and suggestions from Matthew Bershady.
SA would like to thank ESO for the ESO studentship that helped
support part of this work. AAP, KCF, MA, OEG acknowledge the support
of the Australian Research Council Discovery Project grant
DP150104129. AAP acknowledges the support of the STFC consolidated grant
ST/S000488/1, the VICI grant 016.130.338 
of the Netherlands Foundation for Scientific Research (NWO), 
and the Leids KerkhovenBosscha Fonds (LKBF) for travel support. 
The authors are very grateful to the staff at McDonald
Observatory for granting us the observing time and support on the 107
inch telescope. The authors would also like to thank the Isaac Newton
Group staff on La Palma for supporting the PN.S over the years, and
the Swiss National Science Foundation, the Kapteyn Institute, the
University of Nottingham, and INAF for the construction and deployment
of the H$\alpha$ arm. 

\section*{Data availability}
All data used in this paper is available on request to the corresponding authors.

\bibliographystyle{mnras}
\bibstyle{mnras}
\bibliography{N6946}
\vspace*{1cm}

\end{document}